\documentclass[twocolumn,aps,showpacs,superscriptaddress]{revtex4-1}

\usepackage{bm}
\usepackage{amsmath}
\usepackage{amssymb}
\usepackage[usenames,dvipsnames]{color}
\usepackage{graphicx}
\usepackage{dcolumn}
\usepackage{simplewick}
\usepackage{hyperref}
\usepackage{lineno}
\usepackage{float}

\begin{document}

\title{Electric dipole polarizabilities of doubly ionized alkaline Earth 
       metal ions from perturbed relativistic coupled-cluster theory}

\author{S. Chattopadhyay}
\affiliation{Physical Research Laboratory,
             Ahmedabad - 380009, Gujarat,
             India}
\author{B. K. Mani}
\affiliation{Department of Physics, University of South Florida, Tampa,
             Florida 33620, USA}
\author{D. Angom}
\affiliation{Physical Research Laboratory,
             Ahmedabad - 380009, Gujarat,
             India}

\begin{abstract}
  Using perturbed relativistic coupled-cluster (PRCC) theory we compute
  the ground state electric dipole polarizability, $\alpha$, of doubly ionized 
  alkaline earth metal ions  $\rm{Mg}^{2+}$, $\rm{Ca}^{2+}$,  $\rm{Sr}^{2+}$, 
  $\rm{Ba}^{2+}$ and  $\rm{Ra}^{2+}$. In the present work we use the 
  Dirac-Coulomb-Breit atomic Hamiltonian and we also include the Uehling 
  potential, which is the leading order term in the vacuum polarization
  corrections. We examine the correction to the orbital energies arising from 
  the Uehling potential in the self-consistent field calculations as well as 
  perturbatively. Our results of $\alpha$ are in very good agreement with the 
  experimental data, and we observe a change in the nature of the orbital 
  energy corrections arising from the vacuum polarization as we go 
  from $\rm{Mg}^{2+}$ to Ra$^{2+}$. 
\end{abstract}



\maketitle


\section{Introduction}
 
 The static electric dipole polarizability, $\alpha$,  of an atom or ion is 
a measure of the first order response to an external electric field. It is 
an essential parameter to determine any property associated with atom-field
or ion-field interactions as well as atom-atom and atom-ion interactions. 
The properties include the refractive indexes, dielectric constants, ion 
mobility in gases, and van der Waal's constants \cite{bonin-97} and $\alpha$
has been measured using a wide variety of experimental techniques 
\cite{gould-05}. For closed-shell ions, like the doubly ionized 
alkaline-Earth-metal ions, $\alpha$ is a good representative of the 
core-polarization effects. 

 Theoretically, $\alpha$ of the many electron atoms and ions has been 
calculated using different many body methods. A recent review on atomic and 
ionic polarizabilities \cite{mitroy-10} provide a description of the 
theoretical methods used in the calculation of $\alpha$. However, among the
various theoretical methods the ones based on coupled-cluster theory (CCT)
\cite{coester-58, coester-60} are ideal for atoms and ions which are 
closed-shell or with few valence electrons. The CCT is, among the many body 
theories, one of the most reliable and powerful theory. It takes into account 
the electron correlation to all order. A detailed discussion on the CCT 
and different variants are given in a recent review \cite{bartlett-07}, and 
very good descriptions of the application of non-relativistic CCT to atomic
and molecular systems are given in ref. \cite{lindgren-86,shavitt-09}. The 
CCT has been used with great success in atomic 
\cite{mani-09,nataraj-08,pal-07,geetha-01}, molecular 
\cite{isaev-04}, nuclear \cite{hagen-08} and condensed matter physics 
\cite{bishop-09} calculations. For the theoretical calculations of $\alpha$, 
the CCT based methods which have given very precise results are the finite 
field \cite{cohen-65}, sum over states \cite{safronova-99,derevianko-99} and 
perturbed relativistic coupled-cluster (PRCC) theory 
\cite{chattopadhyay-12a, chattopadhyay-12b,chattopadhyay-13}.

  In a previous work, the CCT based finite field method with the 
Douglas-Kroll Hamiltonian \cite{douglas-74} was used to compute the $\alpha$ 
of the alkaline-Earth-metal ions \cite{lim-04}. In this work we compute the 
$\alpha$ of doubly ionized alkaline ions using the PRCC theory. The method
was used in our previous works to calculate the $\alpha$ of noble gas atoms
\cite{chattopadhyay-12a, chattopadhyay-12b} and alkaline-Earth metal ions 
\cite{chattopadhyay-13}. The theory is the conventional relativistic 
coupled-cluster (RCC) theory with an additional perturbation. To account 
for the additional perturbation, we introduce a new set of cluster 
operators and accordingly, define a second set of cluster equations. The
equations, however, are linear in the cluster operators and the new 
operators obey the same selection rules as the perturbation Hamiltonian. In the
calculation of $\alpha$ the perturbation is the external 
electric field $\mathbf{E}$. In the present work we use the Dirac-Coulomb-Breit
atomic Hamiltonian along with the vacuum polarization (VP) potential. The 
VP potential is treated self consistently as well as perturbatively. 

 The paper is organised as follows. In Sec. II we give a brief discussion on
RCC and PRCC theory along with the VP correction. The theoretical formulation
of $\alpha$ in the framework of PRCC theory is discussed in Sec. III. In 
Sec. IV we give the details of our calculational methodology. Next we
discuss about the VP correction to the orbital energies of doubly ionized
alkaline earth metal ions. In the subsequent sections we give the results
of static polarizability and discuss it in great detail. Then we end with
the conclusion. All the results presented in this work and related calculations are in atomic units ( $\hbar=m_e=e=4\pi\epsilon_0=1$). In this system of units 
the velocity of light is $\alpha ^{-1}$, the inverse of fine structure 
constant. For which we use the value of 
$\alpha ^{-1} = 137.035\;999\;074$ \cite{mohr-12}.


\section{Theoretical methods}

  A detailed description of the RCC theory for closed-shell atoms is given in 
ref. \cite{mani-09} and similarly, a detailed account of PRCC theory is 
given in our previous works
\cite{chattopadhyay-12a, chattopadhyay-12b, chattopadhyay-13}. 
However, for completeness and easy reference we provide a brief overview in 
this section.


\subsection{RCC and PRCC theory}
  In the present work we use the Dirac-Coulomb-Breit no-virtual-pair 
Hamiltonian, $H^{\rm DCB}$, to incorporate the relativistic effects and avoid
the difficulties associated with the negative continuum states \cite{brown-51}.
For a doubly ionized atom with $N$ electrons \cite{sucher-80}
\begin{eqnarray}
   H^{\rm DCB} & = & \Lambda_{++}\sum_{i=1}^N \left [c\bm{\alpha}_i \cdot 
        \mathbf{p}_i + (\beta_i -1)c^2 - V_{N+2}(r_i) \right ] 
                       \nonumber \\
    & & + \sum_{i<j}^{N,N}\left [ \frac{1}{r_{ij}}  + g^{\rm B}(r_{ij}) \right ]
        \Lambda_{++},
\end{eqnarray}
where $\bm{\alpha}$ and $\beta$ are the Dirac matrices, $\Lambda_{++}$ is an 
operator which projects to the positive energy solutions and $V_{N+2}(r_{i})$ 
is the nuclear potential arising from the $Z=(N+2)$ nucleus. Projecting the 
Hamiltonian with $\Lambda_{++}$ ensures that the ill effects of the negative 
energy continuum  states are removed from the calculations. An elegant 
alternative to the projection operators, and better suited for numerical
computations, is to use the kinetically balanced finite basis sets
\cite{stanton-84,mohanty-90,grant-06,grant-10}. This is the method adopted in 
the present work to generate the orbital basis sets. Returning to 
$H^{\rm DCB}$, the last two terms, $1/r_{ij} $ and $g^{\rm B}(r_{ij})$,  are 
the Coulomb and Breit interactions, respectively. The later, Breit interaction,
represents the transverse photon interaction and is given by
\begin{equation}
  g^{\rm B}(r_{12})= -\frac{1}{2r_{12}} \left [ \bm{\alpha}_1\cdot\bm{\alpha}_2
               + \frac{(\bm{\alpha_1}\cdot \mathbf{r}_{12})
               (\bm{\alpha_2}\cdot\mathbf{r}_{12})}{r_{12}^2}\right].
\end{equation}
The general trends in the observables arising from the inclusion of Breit 
interaction in RCC and PRCC are discussed in our previous work 
on noble gas atoms \cite{chattopadhyay-12b}. For a closed-shell ion, the 
ground state eigen-value equation is
\begin{equation}
   H^{\rm DCB}|\Psi_0\rangle = E_0|\Psi_0\rangle , 
\end{equation}
where, $|\Psi_0\rangle$ is the ground state of the ion. In the presence of a
perturbation Hamiltonian, $H_{\rm int}$, the eigenvalue equation is modified
to
\begin{equation}
(H^{\rm DCB}+\lambda H_{\rm int})|\tilde{\Psi}_0\rangle 
= \tilde{E}_0|\tilde{\Psi}_0\rangle,
\end{equation}
where $\lambda$ is the perturbation parameter, $|\tilde{\Psi}_0\rangle $ is the
perturbed ground state and $\tilde{E}_0$ is the corresponding eigen energy. 
The origin of $H_{\rm int}$ could be internal to the ion, like the hyperfine 
interaction or external, like the interaction with an external electromagnetic 
field $\mathbf{E}$. 

  In the RCC and PRCC theories, we define two sets of coupled-cluster 
operators $T^{(0)}$ and $\mathbf{T}^{(1)}$, which we refer to as the 
unperturbed and perturbed coupled-cluster operators, respectively. The former
is equivalent to the conventional cluster operators, and the latter 
is an additional set of cluster operators introduced in our recent works
\cite{chattopadhyay-12a,chattopadhyay-12b,chattopadhyay-13}. It accounts for 
the electron correlations effects arising from $H_{\rm int}$ and follows the 
same selection rules as $H_{\rm int}$. To calculate $\alpha$, 
consider the interaction of the ion with an electrostatic electric 
field $\mathbf{E}$. The interaction Hamiltonian is
then
\begin{equation} 
  H_{\rm int} =-\sum_i\mathbf{r}_i\cdot\mathbf{E} = \mathbf{D}\cdot\mathbf{E},
\end{equation}
where $\mathbf{D}$ is the many electron electric dipole operator. The cluster
operators $\mathbf{T^{(1)}}$ are then rank one tensor operators in the 
electronic space and follows the same parity selection rule as $H_{\rm int}$. 
Consequently, as $H_{\rm int}$ is parity odd there is no first order 
perturbative correction to the energy, so to first order in $\lambda$
we get $\tilde{E}_0=E_0$. Using the cluster operators $T^{(0)}$ and
$\mathbf{T}^{(1)}$, the atomic states of unperturbed and perturbed atomic 
Hamiltonians are 
\begin{subequations}
\begin{eqnarray}
  |\Psi_0\rangle  & = & e^{ T^{(0)}}|\Phi_0\rangle , 
  \label{psi_0}                                      \\
  |\tilde{\Psi}_0\rangle & = & e^{\left [ T^{(0)} 
     + \lambda \mathbf{T}^{(1)}\cdot\mathbf{E}\right ]} |\Phi_0\rangle , 
  \label{psi_t}
\end{eqnarray}
\end{subequations}
where $|\Phi_0\rangle$ is the reference state wave-function. The cluster 
operators involve all possible excitations, however, a simplified but accurate
representation is the coupled-cluster single and double (CCSD) excitation
approximation. With this approximation
\begin{subequations}
\begin{eqnarray}
  T^{(0)} & =&  T_{1}^{(0)} + T_{2}^{(0)}, 
  \label{t0_ccsd}   \\
  \mathbf{T}^{(1)} & = & \mathbf{T}_{1}^{(1)} + \mathbf{T}_{2}^{(1)},
  \label{t1_ccsd}
\end{eqnarray}
\end{subequations}
where, the subscripts represent the level of excitation. In the second
quantized notations
\begin{subequations}
\begin{eqnarray}
T_1^{(0)} &= &\sum_{a,p} t_a^p {{a}_p^\dagger} a_a , \\
T_2^{(0)} &= &\frac{1}{2!}\sum_{a,b,p,q}t_{ab}^{pq} {{a}_p^\dagger}{{a}_q^
             \dagger}a_b a_a ,
\end{eqnarray}
\end{subequations}
where $t_{\ldots}^{\ldots}$ are cluster amplitudes, $a_i^{\dagger}$ ($a_i$)
are single particle creation (annihilation) operators and 
$abc\ldots$ ($pqr\ldots$) represent core (virtual) states. Similarly, the
perturbed cluster operators are represented as
\begin{eqnarray}
  \mathbf{T}_1^{(1)} & = & \sum_{a,p} \tau_a^p \mathbf{C}_1 (\hat{r})
                       a_{p}^{\dagger}a_{a},
                           \nonumber \\
  \mathbf{T}_2^{(1)} & = & \sum_{a,b,p,q} \sum_{l,k} \tau_{ab}^{pq}(l,k) 
                   \{ \mathbf{C}_l(\hat{r}_1) \mathbf{C}_k(\hat{r}_2)\}^{1}
                   a_{p}^{\dagger}a_{q}^{\dagger}a_{b}a_{a}. \nonumber
\end{eqnarray}
Here, $\mathbf{C}_1(\hat r)$, a $\mathbf{C}$-tensor is used to represent
the vector nature of $\mathbf{T}_1^{(1)}$. On the other hand, two $\mathbf{C}$ 
tensor operators of rank $l$ and $k$ are coupled together to form a rank one 
tensor operator, $\mathbf{T}_2^{(1)}$.  For a more rigorous description of
the tensor structure of the PRCC operators we refer to our previous work
\cite{chattopadhyay-12b}.


\subsection{Vacuum Polarization}
In the present work we incorporate the vacuum polarization (VP) 
corrections to the electron-nucleus interactions. It modifies the Coulomb 
potential between the nucleus and electrons. For a point nucleus, to the 
order of $Z\alpha$, it is given by the Uehling potential\cite{uehling-35}
\begin{equation}
   V_{\rm Ue}(r) = -\frac{2\alpha Z}{3\pi r}
                    \int_{1}^{\infty} dt\sqrt{t^2 - 1}\big(\frac{1}{t^2} + 
                    \frac{1}{2t^4}\big) \exp{[-\frac{2rt}{\alpha}]}, \nonumber
\end{equation}
where $Z$ is the nuclear charge and $\alpha$, in this case, is the fine
structure constant. The latter is not to be confused with the dipole 
polarizability. In heavy atoms a finite size Fermi charge distribution model 
of the nucleus is more appropriate \cite{parpia-92a} and it is 
defined as
\begin{equation}
   \rho_{\rm nuc}(r) = \frac{\rho_0}{1 + e^{(r-c)/a}},
\end{equation}
here $a=t 4\ln(3)$. The parameter $c$ is the half charge radius so that
$\rho_{\rm nuc}(c) = {\rho_0}/{2}$ and $t$ is the skin thickness. For a 
consistent treatment of the nucleus-electron electrostatic interaction, 
$V_{\rm Ue}(r)$ must be modified to account for the finite nuclear size. This
is done by folding $V_{\rm Ue}(r)$ with the $\rho_{\rm nuc}(r)$
\cite{fullerton-76}. The
modified form of the Uehling potential is \cite{johnson-01}
\begin{eqnarray}
  {V}_{\rm Ue}{(r)} = -\frac{2\alpha^{2}}{3r}
                       \int_{0}^{\infty}\mathrm {d}{\rm x} \, x \rho{(x)}
                       \int_{1}^{\infty}\mathrm{d}t \sqrt{t^{2}-1} \nonumber \\ 
                       \Big( {\frac{1}{t^{3}} + \frac{1}{2t^{5}}} \Big) 
                       (e^{-2ct|(r-x)|} - e^{-2ct(r+x)}). \nonumber
\end{eqnarray}
We add this to the electron-nucleus Coulomb interaction potential in the 
self-consistent field computations to generate the single particle states. 
The Uehling potential is the leading order term in VP correction and it 
accounts for more than 90\% of the VP correction in Hydrogen like ions. So we 
identify it as the VP correction in the subsequent sections.


\subsection{Linearized PRCC Theory}
In this section we describe in brief the linearized form of the PRCC (LPRCC)
theory. It is much simpler than the complete PRCC but encompasses all the 
important many-body effects. To derive the LPRCC equations, as discussed
earlier, consider $\mathbf{E}$ as the perturbation. The eigen-value 
equation is then
\begin{equation}
  (H^{\rm DCB} + \lambda H_{\rm int}) e^{\left [T^{(0)} + \lambda 
  \mathbf{T}^{(1)}\cdot\mathbf{E} \right ]} |\Phi_0\rangle 
  = \tilde{E}_{0} e^{\left [T^{(0)} 
  + \lambda \mathbf{T}^{(1)}\cdot\mathbf{E}\right ]} |\Phi_0\rangle .
 \label{prcc_eival}
\end{equation}
Using the normal-ordered form of Hamiltonian the eigen value equation may be
written as
\begin{equation}
  \left (H^{\rm DCB}_{\rm N} + \lambda H_{\rm int}\right )|\tilde{\Psi}_0\rangle
   = \Delta E_0|\tilde{\Psi}_0\rangle ,
\end{equation}
where, $\Delta E_0 = E_0 - \langle \Phi_0|H^{\rm DCB}|\Phi_0\rangle$ is the
ground state correlation energy of the many-electron ion. Using the PRCC
wave-function in Eq. (\ref{psi_t}), we write the ground state as
\begin{eqnarray}
 |\tilde{\Psi}_0\rangle \approx e^{T^{(0)}}\left [ 1 
    + \lambda \mathbf{T^{(1)}\cdot \mathbf{E}} \right ] |\Phi_0\rangle .
 \label{psi_tilde}
\end{eqnarray}
Using this expression, the PRCC eigen-value equation assumes the form
\begin{eqnarray}
  && \left (H^{\rm DCB}_{\rm N} + \lambda H_{\rm int}\right )e^{T^{(0)}}
    \left [ ( 1 + \lambda \mathbf{T}^{(1)}\cdot\mathbf{E})\right ] 
    |\Phi_0\rangle = \nonumber \\
  &&  \Delta E_0e^{T^{(0)}}\left [( 1 + \lambda \mathbf{T}^{(1)}\cdot
    \mathbf{E})\right ] |\Phi_0\rangle .
\end{eqnarray}
Following the standard coupled-cluster ansatz, as the initial step to derive 
the cluster amplitude equations, we apply $e^{-T^{(0)}}$ from the left and get
\begin{equation}
  \left [\bar H^{\rm DCB}_{\rm N}  + \lambda \bar H_{\rm int}\right ] 
    e^{\lambda \mathbf{T}^{(1)}\cdot\mathbf{E}}|\Phi_0\rangle 
    = \Delta E_0e^{\lambda \mathbf{T}^{(1)}\cdot\mathbf{E}} |\Phi_0\rangle ,
  \label{prcc_eq1}
\end{equation}
where $\bar{H} = e^{-T^{(0)}}H e^{T^{(0)}}$ is the similarity transformed 
Hamiltonian. After applying $e^{-\lambda \mathbf{T}^{(1)}}$ from the left
and considering the terms linear in $\lambda$, we obtain the 
PRCC equation
\begin{equation}
   \left ( \left [\bar{H}^{\rm DC}_{\rm N},\mathbf{T}^{(1)}\right ]
      \cdot\mathbf{E} + \bar{H}_{\rm int}\right ) |\Phi_0\rangle = 0 .
\end{equation}
The linearized PRCC is the approximation where we take
$   \left [\bar{H}^{\rm DC}_{\rm N},\mathbf{T}^{(1)}\right ] 
\approx \left [H^{\rm DC}_{\rm N},\mathbf{T}^{(1)}\right ]$ and
$
\bar{H}_{\rm int} \approx \mathbf{D} + \left[\mathbf{D},T^{(0)}\right ]
$. The eigenvalue equation is then reduced to
\begin{equation}
  \left [H_{\rm N}^{\rm DCB},\mathbf{T}^{(1)}\right ] |\Phi_0\rangle
      =  \bigg ( \mathbf{D} + \left [\mathbf{D},T^{(0)}\right ]  
     \bigg )|\Phi_0\rangle . \;\;\;
   \label{prcc_eq2}
\end{equation}
Where, for simplicity, we have dropped $\mathbf{E}$ from the equation.
The equations of the cluster amplitudes  $\mathbf{T}^{(1)}_1$
and $\mathbf{T}^{(1)}_2$ are obtained by projecting the above equation to
single and double excited states $\langle \Phi_a^p|$ and 
$\langle\Phi_{ab}^{pq}| $, respectively. These states, however, must be 
opposite in parity to the reference state $|\Phi_0\rangle $. The equations
so obtained forms a set of linear algebraic equations and are solved using 
standard linear algebraic methods.

 The other method of calculating $\alpha$ which avoids summation over the
intermediate states is the finite field method \cite{cohen-65}. The method,
however, requires evaluation of the energy for different values of 
$\mathbf{E}$ and this implies computing the cluster amplitudes multiple times. 
In the PRCC theory, however, the computations of the cluster amplitudes are 
limited to one time evaluation of ${T}^{(0)} $  and $\mathbf{T}^{(1)}$. 
Although, the equations of $\mathbf{T}^{(1)}$ are linear, the tensor nature 
translates into angular factors consisting of a large numbers of $6j$-symbols 
and $9j$-symbols. So, for our present work we resort to a symmetry adapted 
storing of these angular factors.


\section{Dipole Polarizability}
In the present calculation of $\alpha$ we use the PRCC expression 
discussed and described in our previous works 
\cite{chattopadhyay-12a, chattopadhyay-12b}.  Accordingly, the $\alpha$ of the 
ground state of a doubly ionized alkaline atom is
\begin{equation}
  \alpha = -\frac{\langle \Phi_0|\mathbf{T}^{(1)\dagger}\bar{\mathbf{D}} + 
   \bar{\mathbf{D}}\mathbf{T}^{(1)}|\Phi_0\rangle}{\langle\Psi_0|\Psi_0\rangle},
\end{equation}
Here $\bar{\mathbf{D}} = e^{{T}^{(0)\dagger}}\mathbf{D} e^{T^{(0)}}$ is a
non-terminating series and we have consider only the leading order terms in 
this expression.   
\begin{eqnarray}
 \alpha &=& -\frac{1}{\cal N}\langle\Phi_0|\mathbf{T}_1^{(1)\dagger}\mathbf{D} 
       + \mathbf{D}\mathbf{T}_1^{(1)} 
       + \mathbf{T}_1^{(1)\dagger}\mathbf{D}T_2^{(0)}
       + T_2^{(0)\dagger}\mathbf{D}\mathbf{T}_1^{(1)} \nonumber \\
    && + \mathbf{T}_1^{(1)\dagger}\mathbf{D}T_1^{(0)} 
       + T_1^{(0)\dagger}\mathbf{D}\mathbf{T}_1^{(1)}
       + \mathbf{T}_2^{(1)\dagger}\mathbf{D}T_1^{(0)} \nonumber \\
    && + T_1^{(0)\dagger}\mathbf{D}\mathbf{T}_2^{(1)}
       + \mathbf{T}_2^{(1)\dagger}\mathbf{D}T_2^{(0)} 
       + T_2^{(0)\dagger}\mathbf{D}\mathbf{T}_2^{(1)}
     |\Phi_0\rangle, 
  \label{dip_pol_eqn}
\end{eqnarray}
where ${\cal N} = \langle\Phi_0|\exp[T^{(0)\dagger}]\exp[T^{(0)}]
|\Phi_0\rangle$ is the normalization factor, which involves a non-terminating
series of contractions between ${T^{(0)}}^\dagger $ and $T^{(0)} $. However, in 
the present work we use 
${\cal N} \approx \langle\Phi_0|1 + T_1^{(0)\dagger}T_1^{(0)} + 
T_2^{(0)\dagger}T_2^{(0)}|\Phi_0\rangle$. In the PRCC expression of $\alpha$,
the summation over intermediate states is subsumed within 
$\mathbf{T}^{(1)}$ in a natural way and eliminates the need for 
a complete set of intermediate states. This is, however, with the condition 
of solving an additional set of cluster equations.


\section{Calculational details}
\subsection{Basis set}

 To get accurate results the first step is to generate an appropriate basis set 
of orbitals. Here we use the Gaussian type orbitals (GTO's), in which the
orbitals are expressed as a linear combination of Gaussian type functions
\cite{mohanty-90}. In particular, the large component of the orbitals
are the linear combination of the Gaussian type functions of the form
\begin{equation}
   g_{\kappa p}^{L}(r) = C^{L}_{\kappa i} r^{n_{\kappa}}e^{-\alpha_{p}r^{2}},
\end{equation}
where $p=0,1\ldots m$ is the GTO index and $m$ is the number of Gaussian type
functions. The exponent $\alpha_{p} = \alpha_{0} \beta^{p-1}$, where
$\alpha_{0}$ and $\beta$ are two independent parameters. The small component
are constructed from the large component through the kinetic balance
condition\cite{stanton-84,mohanty-90,grant-06,grant-10}. The GTOs are 
calculated on a grid \cite{chaudhuri-99} and we optimize the values of 
$\alpha_{0}$ and $\beta$ for individual atoms to reproduce the orbital 
energies of the core orbitals and self consistent field (SCF) energy from 
GRASP92 \cite{parpia-96} code. The comparison of the SCF energies for the 
doubly ionized alkaline atoms are given in Table. \ref{orb}. 
\begin{table}[h]
        \caption{Comparison between the ground state SCF energies obtained
                 from the computations with GTO and GRASP92. The energies are
                 in atomic units.}
        \label{orb}
        \begin{center}
        \begin{tabular}{ldd}
            \hline
            Atom &
              \multicolumn{1}{c}{GTO} &
              \multicolumn{1}{c}{GRASP92} \\
            \hline
            $\rm{Mg}^{2+}$ & -199.1500   & -199.1501   \\
            $\rm{Ca}^{2+}$ & -679.1038   & -679.1038   \\
            $\rm{Sr}^{2+}$ & -3177.5211  & -3177.5218  \\
            $\rm{Ba}^{2+}$ & -8135.1404  & -8135.1421  \\
            $\rm{Ra}^{2+}$ & -26027.5632 & -26027.5634 \\
           \hline
        \end{tabular}
        \end{center}
\end{table}
From the table it is evident that the results of the SCF energies from the 
GTOs are in agreement with the GRASP92 results to the accuracy of at least
$10^{-3}$ Hartree. The symmetry wise values of the optimized 
$\alpha_{0}$ and $\beta$ are listed in Table. \ref{basis}  
\begin{table}[h]
   \caption{The $\alpha_0$ and $\beta$ parameters of the even tempered
            GTO basis for different ions used in the present calculations.}
   \label{basis}
   \begin{tabular}{cccccccc}
   \hline
   \hline
     Atom & \multicolumn{2}{c}{$s$} & \multicolumn{2}{c}{$p$} &
     \multicolumn{2}{c}{$d$}  \\
     & $\alpha_{0}$  & $\beta$ & $\alpha_{0}$ & $\beta$
     & $\alpha_{0}$  & $\beta$  \\
     \hline
     $\rm{Mg}^{2+}$ &\, 0.00825  &\, 2.310 &\, 0.00715 &\, 2.365 &\, 0.00700
     &\, 2.700 \\
     $\rm{Ca}^{2+}$ &\, 0.00895  &\, 2.110 &\, 0.00815 &\, 2.150 &\, 0.00750
     &\, 2.500 \\
     $\rm{Sr}^{2+}$ &\, 0.00975  &\, 2.100 &\, 0.00915 &\, 2.010 &\, 0.00900
     &\, 2.030 \\
     $\rm{Ba}^{2+}$ &\, 0.00985  &\, 2.150 &\, 0.00975 &\, 2.070 &\, 0.00995
     &\, 2.010 \\
     $\rm{Ra}^{2+}$ &\, 0.00995  &\, 2.110 &\, 0.00925 &\, 2.090 &\, 0.00850
     &\, 2.010 \\
     \hline
   \end{tabular}
\end{table}

 The number of Gaussian type functions with the optimized basis set parameters 
is large and not all the GTOs generated are important for the calculations.
For the PRCC calculation we select the number of GTO's for each symmetry such 
that the electron correlation is accounted accurately. In order to investigate 
this, we examine the convergence pattern of the $\alpha$ by varying the basis 
set size. Here we present the result for $\rm{Sr}^{2+}$. We start with a 
basis set size of 95 GTOs and increase it in steps upto 155 GTO's. For this
the computations are done with the Dirac-Coulomb Hamiltonian and the results
are listed in table \ref{basis_sr}. Based on the table the optimal basis size
to get converged result accurate upto $10^{-3}$ is 127.
\begin{table}[h]
  \caption{Convergence pattern of $\alpha$ of $\rm{Sr}^{2+}$ ion as a function 
           of the basis set size. For this set of calculations we consider the
           Dirac-Coulomb Hamiltonian and result is in atomic units.}
  \label{basis_sr}
  \begin{tabular}{lcc}
      \hline
      No. of orbitals & Basis size & $\alpha $   \\
      \hline
      95   & $(15s, 11p, 11d,   9f,   9g)  $ &  5.762   \\
      113  & $(17s, 13p, 13d,  11f,  11g)  $ &  5.745   \\
      127  & $(19s, 15p, 15d,  13f,  11g)  $ &  5.743   \\
      137  & $(21s, 17p, 17d,  13f,  11g)  $ &  5.743   \\
      155  & $(23s, 19p, 19d,  15f,  13g)  $ &  5.743   \\
      \hline
  \end{tabular}
\end{table}

To solve the PRCC equations for single and double excitations,
we use Jacobi method. We chose this method as it can be parallelized 
without any difficulty. However, there is a major drawback of the method or
performance penalty: slow convergence. To accelerate the convergence we use 
direct inversion of the iterated subspace(DIIS) \cite{pulay-80} and this 
improves the convergence significantly.


\subsection{VP Corrections to the Orbital Energies}

To study the VP corrections arising from $V_{\rm Ue}$, we compute the 
orbital energy corrections in the self consistent field (SCF) calculations. We 
also compute the first order correction using the many-body perturbative 
theory. In the former case, SCF calculations, the VP potential is considered 
along with the Dirac-Hartree-Fock (DHF) potential, $U_{\rm DHF}$. The 
orbital eigen-value equation is then
\begin{equation}
   \left [ h_{0} + V_{\rm Ue}(r) + U_{\rm DHF}(r) \right] |\psi'_i \rangle = 
   \epsilon'_i |\psi '_i \rangle,  \nonumber  
\end{equation}
where, $h_0 = c\bm{\alpha}\cdot\mathbf{p} +(\beta - 1)c^2 -V_{\rm N-2}(r)$ is
the single particle part of Dirac-Coulomb Hamiltonian, $U_{\rm DHF}(r)$ is
the Dirac-Hartree-Fock potential,   $|\psi'_i \rangle$ is a four component orbital and $\epsilon'_i$ is the
corresponding eigenvalue. Similarly, we use {\em unprimed} states, 
$|\psi_i\rangle$, to represent orbitals which are eigenfunctions of the DHF 
Hamiltonian, that is
\begin{equation}
   \left [ h_{0} + U_{\rm DHF}(r) \right ] |\psi_i \rangle = 
   \epsilon_i |\psi_i \rangle,  \nonumber  
\end{equation}
where $\epsilon_i $ is the DHF energy of the orbital. To quantify the VP effect
we define 
\begin{equation}
   \Delta\epsilon_{i} = \epsilon' _i - \epsilon_i,
\end{equation}
as the change in the orbital energy due to $V_{\rm Ue}(r)$. Following the
time-independent many-body perturbation theory, the first order energy
correction associated with  $V_{\rm Ue}(r)$ is
\begin{equation}
   \langle V_{\rm Ue}\rangle_i = \langle \psi_i|V_{\rm Ue}(r)|\psi_i \rangle.
     \nonumber  
\end{equation}
Since the VP potential is attractive and short range in nature, it has 
larger effect on the orbitals which have finite probability density within 
the nucleus. So, at the first order $\langle V_{\rm Ue}\rangle$  is negative 
for orbitals, but only the $s_{1/2}$ orbitals have negative $\Delta\epsilon$ 
for all the ions. A similar pattern is reported in ref. \cite{derevianko-04} 
for the orbitals energies of $\rm{Cs}^{+}$. For the Ra$^{2+}$ ion, in addition 
to $s_{1/2}$ the $p_{1/2}$  orbitals also have negative $\Delta\epsilon$. 
More details of the $\Delta\epsilon_{i}$ and $\langle V_{\rm Ue}\rangle_i $ 
for the core orbitals of the $\rm{Ca}^{2+}$, $\rm{Sr}^{2+}$, $\rm{Ba}^{2+}$ 
and $\rm{Ra}^{2+}$, and are presented in the next section.

\begin{table}[h]
        \caption{Static dipole polarizability of doubly ionized 
                 alkaline-Earth-metal  ions and the values are in atomic
                 units. }
        \label{pol_alkaline}
        \begin{center}
        \begin{tabular}{ldcdc}
            \hline
            \multicolumn{1}{c}{Atom} & \multicolumn{1}{c}{This Work} & 
            \multicolumn{1}{c}{Method}                               & 
            \multicolumn{1}{c}{Previous Works}                       &
            \multicolumn{1}{c}{Method}                              \\  
            \hline
            \hline
            $\rm{Mg}^{2+}$ & 0.489  & (LPRCC) &  0.469   \footnotemark[1] &  
            RRPA   \\
                           & 0.495  & (PRCC)  &  0.489(5)\footnotemark[2] & 
            Expt.  \\
            &&&& \\ 
            $\rm{Ca}^{2+}$ & 3.284  & (LPRCC) &  3.262   \footnotemark[3] & 
            RCCSDT \\
                           & 3.387  & (PRCC)  &  3.254   \footnotemark[1] &                 RRPA   \\
                           &        &         &  3.26(3) \footnotemark[2] &                 Expt.  \\
            &&&& \\ 
            $\rm{Sr}^{2+}$ & 5.748  & (LPRCC) &  5.792   \footnotemark[3] & 
            RCCSDT \\
                           & 5.913  & (PRCC)  &  5.813   \footnotemark[1] &                 RRPA   \\
            &&&& \\ 
            $\rm{Ba}^{2+}$ & 10.043 & (LPRCC) & 10.491   \footnotemark[3] & 
            RCCSDT \\
                           & 10.426 & (PRCC)  & 10.61    \footnotemark[1] &
            RRPA   \\
            &&&& \\ 
            $\rm{Ra}^{2+}$ & 12.908 & (LPRCC) & 13.361   \footnotemark[3] &                 RCCSDT \\
                           & 13.402 & (PRCC)  &        &                 \\
            \hline
            \hline
        \end{tabular}
        \end{center}
\footnotetext[1]{Reference\cite{johnson-83}.}
\footnotetext[2]{Reference\cite{opik-67}.}
\footnotetext[3]{Reference\cite{lim-04}.}
\end{table}


\section{Results and Discussions}
  As mentioned earlier, the expression of the $\alpha$ in PRCC theory 
is a non-terminating series of the cluster amplitudes. However, considering 
that the cluster operators $T_2^{(0)}$ and $\mathbf{T}_1^{(1)}$ accounts for 
more than 95\% of the many-body effects in RCC and PRCC, the terms considered 
in Eq. (\ref{dip_pol_eqn}) give very accurate results. To verify, we have 
examined the leading terms which are third order in cluster amplitudes and 
find the contributions are $\sim 10^{-4}$. So, for the present work, as
we consider $\alpha$ upto third decimal place, it is appropriate to neglect 
the contributions from terms which are third and higher order in cluster 
operators. 

  In table \ref{pol_alkaline} we list the $\alpha$ of alkaline-Earth metal 
ions Mg$^{2+}$, Ca$^{2+}$, Sr$^{2+}$, Ba$^{2+}$ and Ra$^{2+}$ 
computed using Eq. (\ref{dip_pol_eqn}). The results are based on two sets of 
calculations: one is based on the cluster amplitudes obtained from LPRCC and 
the other is based on PRCC. For a systematic comparison we also list the 
previous theoretical and experimental results. The results of $\alpha$ along
with the orbital energy corrections arising from $V_{\rm Ue}(r) $ 
for each of the ions are discussed in the subsequent sections.


\subsection{$\rm{Mg}^{2+}$}
The $\alpha$ of $\rm{Mg}^{2+}$  computed with LPRCC is in excellent agreement 
with the experimental data. However, the PRCC result is 1.2\% higher than the
LPRCC result and experimental data. This may be due to a part of the 
additional many-body effects arising from the nonlinear terms in the cluster 
amplitude equations, but which may ultimately cancel with the contributions 
from the cluster amplitudes of higher excitations like $ T_3^{(0)}$ and
$\mathbf{T}_3^{(1)}$. The RRPA result is 4.1\% lower than the experimental 
data and it is also lower than both the LPRCC and PRCC results. It must be 
added that a similar trend is observed for the Na$^+$ ion 
\cite{chattopadhyay-13}, which is isoelectronic with $\rm{Mg}^{2+}$, the RRPA 
result of $\alpha$ is lower than the experimental data \cite{johnson-83}. 
This trend may be on account of the inherent strength and limitation of RRPA, 
the potential to incorporate core-polarization effects very accurately and 
weakness to account for pair correlation effects.

To estimate the contribution from the Breit interaction we consider the 
Dirac-Coulomb Hamiltonian with the VP potential. The contribution from the
Breit interaction can be safely neglected for this ion as the contribution is 
less than 0.02\%. Not surprisingly, the orbital energy corrections 
$\Delta\epsilon_{i}$  and $\langle V_{\rm Ue}\rangle_i$ are very small and can
be neglected. For this reason we have not listed the values of 
$\Delta\epsilon_{i}$  and $\langle V_{\rm Ue}\rangle_i$ for Mg$^{2+}$.
\begin{table}[h]
   \caption{VP Corrections to the orbital energies of
            $\rm{Ca}^{2+}$. Here [x] represents multiplication by ${10^x}$.}
   \label{vp_ca}
   \begin{tabular}{ldd}
   \hline
   \hline
     Orbital & \multicolumn{1}{c}{\;\;\;\;\;\;$\Delta\epsilon$} & 
     \multicolumn{1}{c}{\;\;\;\;\;\;$\langle V_{\rm Ue}\rangle$}  \\
     \hline
     $1s_{1/2}$ &\, -4.204[-3]  &\, -4.435[-3] \\
     $2s_{1/2}$ &\, -3.531[-4]  &\, -3.790[-4] \\
     $2p_{1/2}$ &\,  4.884[-5]  &\, -1.511[-6] \\
     $2p_{3/2}$ &\,  4.938[-5]  &\, -2.732[-7] \\
     $3s_{1/2}$ &\, -4.391[-5]  &\, -4.500[-5] \\
     $3p_{1/2}$ &\,  6.817[-6]  &\, -1.619[-7] \\
     $3p_{3/2}$ &\,  6.880[-6]  &\, -2.931[-8] \\
     \hline
   \end{tabular}
\end{table}


\subsection{$\rm{Ca}^{2+}$}
For $\rm{Ca}^{2+}$, the LPRCC result of $\alpha$ is within the experimental
uncertainty and it is in good agreement with the result from a previous work,
which is based on the RCCSDT theory. The PRCC result is 3.1\% larger than 
the LPRCC result and deviates from the experimental data by 3.7\%. On the 
other hand, the result from the RRPA \cite{johnson-83}, like in Mg$^{2+}$, is 
lower than the experimental data. 

  Based on another set of calculations with the Dirac-Coulomb Hamiltonian,
the contribution from the Breit interaction is estimated to be 
0.004, which is a mere $\approx$0.1\% of the total value. Similarly, we 
calculate the VP correction to the orbital energy with a series of SCF
calculations and results are listed in Table. \ref{vp_ca}. As to be expected, 
the first order correction $\langle V_{\rm Ue}\rangle $ is negative for all 
the core orbitals. But the values of $\Delta\epsilon$ are negative only for 
the $s_{1/2}$ orbitals. Another important observation is, for $s_{1/2}$ orbitals
$\langle V_{\rm Ue}\rangle_i$ and $\Delta\epsilon_{i}$ are similar in value.
But for the other orbitals, besides the change in sign, the values of
$\langle V_{\rm Ue}\rangle_i$ and $\Delta\epsilon_{i}$ are different by several
orders of magnitude.
\begin{table}[h]
   \caption{VP Corrections to the orbital energies of $\rm{Sr}^{2+}$. 
            Here [x] represents multiplication by ${10^x}$.}
   \label{vp_sr}
   \begin{tabular}{ldd}
   \hline
   \hline
     Orbital & \multicolumn{1}{c}{\;\;\;\;\;\;$\Delta\epsilon$} & 
     \multicolumn{1}{c}{\;\;\;\;\;\;$\langle V_{\rm Ue}\rangle$}    \\
     \hline
     $1s_{1/2}$ &\,  -5.721[-2]  &\, -5.904[-2] \\
     $2s_{1/2}$ &\,  -5.968[-3]  &\, -6.231[-3] \\
     $2p_{1/2}$ &\,   3.604[-4]  &\, -1.144[-4] \\
     $2p_{3/2}$ &\,   4.354[-4]  &\, -1.636[-5] \\
     $3s_{1/2}$ &\,  -1.003[-3]  &\, -1.045[-3] \\
     $3p_{1/2}$ &\,   8.281[-5]  &\, -1.995[-5] \\
     $3p_{3/2}$ &\,   9.664[-5]  &\, -2.865[-6] \\
     $3d_{3/2}$ &\,   8.145[-5]  &\, -4.341[-9] \\
     $3d_{5/2}$ &\,   8.048[-5]  &\, -1.123[-9] \\
     $4s_{1/2}$ &\,  -1.301[-4]  &\, -1.320[-4] \\
     $4p_{1/2}$ &\,   1.592[-5]  &\, -2.086[-6] \\
     $4p_{3/2}$ &\,   1.747[-5]  &\, -2.984[-7] \\
     \hline
   \end{tabular}
\end{table}


\subsection{$\rm{Sr}^{2+}$} 
 For $\rm{Sr}^{2+}$ it is important to have accurate theoretical results as
there are no experimental data of $\alpha$. From the Table \ref{pol_alkaline} 
the LPRCC result of 5.748 is  in very good agreement with the previous work 
using RCCSDT. And, like in the previous cases, the PRCC result of 5.913 is 
larger than the LPRCC result. Comparing the results from different 
theoretical methods, we observe the emergence of two important changes in the 
relative patterns when compared with the results results of Mg$^{2+}$ and 
Ca$^{2+}$.  First, the RRPA result is higher than both the LPRCC and RCCSDT 
results, and second, the RCCSDT result is larger than the LPRCC result. This 
may be on account of the filled $3d$ shell in Sr$^{2+}$. As it is of higher 
angular momentum, it has larger polarization effects as well as pair 
correlation effects. A method like RRPA incorporates the core-polarization 
effects very accurately but could potentially under estimate the pair 
correlation effects. Not surprisingly, the same trends are observed in the 
heavier ions Ba$^{2+}$ and Ra$^{2+}$ with filled $d$ and $f$ shells.

Based on a comparison with the calculations using the Dirac-Coulomb 
Hamiltonian, we estimate the Breit contribution as 0.005. This is negligibly 
small and similar in magnitude to the case of Ca$^{2+}$. The VP corrections to 
the orbital energies arising from $V_{\rm Ue}(r)$ are listed in 
Table. \ref{vp_sr}. From the table it is evident that 
$\Delta\epsilon_{1s_{1/2}}$ is an order of magnitude larger than in 
Ca$^{2+}$. In addition, we also observe a four orders of magnitude difference 
between the $\langle V_{\rm Ue}\rangle_i$ and $\Delta\epsilon_{i}$ of the
$3d$ orbitals. This is not surprising as the short range $V_{\rm Ue}(r)$ 
have little effect on the electrons in the higher angular momentum orbitals 
like $d$.
\begin{table}[h]
   \caption{VP Corrections to the orbital energies of $\rm{Ba}^{2+}$. 
            Here [x] represents multiplication by ${10^x}$.}
   \label{vp_ba}
   \begin{tabular}{ldd}
   \hline
   \hline
     Orbital & \multicolumn{1}{c}{\;\;\;\;\;\;$\Delta\epsilon $} & 
     \multicolumn{1}{c}{\;\;\;\;\;\;$\langle V_{\rm Ue}\rangle$}  \\
     \hline
     $1s_{1/2}$ &\, -2.952[-1]  &\, -3.025[-1] \\
     $2s_{1/2}$ &\, -3.493[-2]  &\, -3.623[-2] \\
     $2p_{1/2}$ &\,  5.074[-4]  &\, -1.669[-3] \\
     $2p_{3/2}$ &\,  1.786[-3]  &\, -1.748[-4] \\
     $3s_{1/2}$ &\, -7.084[-3]  &\, -7.391[-3] \\
     $3p_{1/2}$ &\,  1.984[-4]  &\, -3.725[-4] \\
     $3p_{3/2}$ &\,  4.926[-4]  &\, -3.981[-5] \\
     $3d_{3/2}$ &\,  4.856[-4]  &\, -2.047[-7] \\
     $3d_{5/2}$ &\,  4.737[-4]  &\, -4.712[-8] \\
     $4s_{1/2}$ &\, -1.531[-3]  &\, -1.599[-3] \\
     $4p_{1/2}$ &\,  8.513[-5]  &\, -7.689[-5] \\
     $4p_{3/2}$ &\,  1.476[-4]  &\, -8.242[-6] \\
     $4d_{3/2}$ &\,  1.272[-4]  &\, -4.004[-8] \\
     $4d_{5/2}$ &\,  1.245[-4]  &\, -9.185[-8] \\
     $5s_{1/2}$ &\, -2.449[-4]  &\, -2.473[-4] \\
     $5p_{1/2}$ &\,  2.295[-5]  &\, -1.071[-5] \\
     $5p_{3/2}$ &\,  3.230[-5]  &\, -1.066[-6] \\
     \hline
   \end{tabular}
\end{table}


\subsection{$\rm{Ba}^{2+}$}
Like in Sr$^{2+}$, there are no experimental data of $\alpha$ for Ba$^{2+}$. 
Hence, it is important to have accurate theoretical results and in this regard,
it is pertinent to calculate $\alpha$ with a reliable method like RCC. Here,
computing with the relativistic version coupled-cluster is essential as the 
high $Z$ implies that the relativistic corrections are important. From 
Table. \ref{pol_alkaline}, it is evident that our LPRCC result of
10.043 is 4.3\% lower than the RCCSDT result. However, our PRCC result
is in very good agreement with the RCCSDT result, it is just 0.6\% less. 
Examining the results discussed so far, there is a discernible trend when
we compare the PRCC and RCCSDT results. The difference between the two
results narrows with increasing $Z$. This may be due to the the basic property 
of the CCT, the inclusion of selected electron correlation effects to all 
order. So, with higher $Z$ the importance of the correlation effects grows 
and the two coupled-cluster based methods incorporate the correlation effects 
to similar accuracy. The other theoretical result from the RRPA theory is 
larger than the other results.

 Following the computations described earlier, we estimate the 
Breit contribution as 0.007, which is similar to the previous cases. Coming 
to the orbital energy corrections arising from the VP, we find an important
change in the pattern of $\Delta\epsilon$. The $\Delta\epsilon$ of $p_{1/2}$
and $p_{3/2}$ continue to be positive, but $\Delta\epsilon_{2p_{1/2}}$ is 
$\approx$72\% smaller than $\Delta\epsilon_{2p_{3/2}}$. For the remaining
$np_{1/2}$ and $np_{3/2}$, although the difference is not so dramatic, the
differences are still large. 
\begin{table}[h]
   \caption{VP Corrections to the orbital energies of $\rm{Ra}^{2+}$. 
            Here [x] represents multiplication by ${10^x}$.}
   \label{vp_ra}
   \begin{tabular}{ldd}
   \hline
   \hline
     Orbital & \multicolumn{1}{c}{\;\;\;\;\;\;$\Delta\epsilon$} & 
     \multicolumn{1}{c}{\;\;\;\;\;\;$\langle V_{\rm Ue}\rangle$}  \\
     \hline
     $1s_{1/2}$ &\, -2.560      &\, -2.614     \\
     $2s_{1/2}$ &\, -3.881[-1]  &\, -3.999[-1] \\
     $2p_{1/2}$ &\, -3.802[-2]  &\, -5.753[-2] \\
     $2p_{3/2}$ &\,  1.211[-2]  &\, -2.707[-3] \\
     $3s_{1/2}$ &\, -8.999[-2]  &\, -9.315[-2] \\
     $3p_{1/2}$ &\, -9.620[-3]  &\, -1.504[-2] \\
     $3p_{3/2}$ &\,  3.728[-3]  &\, -7.545[-4] \\
     $3d_{3/2}$ &\,  4.213[-3]  &\, -1.330[-5] \\
     $3d_{5/2}$ &\,  3.953[-3]  &\, -2.385[-6] \\
     $4s_{1/2}$ &\, -2.362[-2]  &\, -2.451[-2] \\
     $4p_{1/2}$ &\, -2.238[-3]  &\, -3.938[-3] \\
     $4p_{3/2}$ &\,  1.315[-3]  &\, -1.999[-4] \\
     $4d_{3/2}$ &\,  1.350[-3]  &\, -3.943[-6] \\
     $4d_{5/2}$ &\,  1.282[-3]  &\, -7.062[-7] \\
     $4f_{5/2}$ &\,  1.015[-3]  &\, -1.647[-9] \\
     $4f_{7/2}$ &\,  9.928[-4]  &\, -4.229[-10] \\
     $5s_{1/2}$ &\, -5.378[-3]  &\, -5.633[-3] \\
     $5p_{1/2}$ &\, -3.002[-4]  &\, -8.438[-4] \\
     $5p_{3/2}$ &\,  4.845[-4]  &\, -4.200[-5] \\
     $5d_{3/2}$ &\,  4.074[-4]  &\, -6.735[-7] \\
     $5d_{5/2}$ &\,  3.859[-4]  &\, -1.187[-7] \\
     $6s_{1/2}$ &\, -9.883[-4]  &\, -9.951[-4] \\
     $6p_{1/2}$ &\, -1.613[-5]  &\, -1.290[-4] \\
     $6p_{3/2}$ &\,  1.211[-4]  &\, -5.949[-6] \\
     \hline
   \end{tabular}
\end{table}


\subsection{$\rm{Ra}^{2+}$}
Our PRCC result of $\alpha$ for Ra$^{2+}$ is $\approx$3.7\% larger than the 
LPRCC result. This trend is similar to the case of Ba$^{2+}$ and may be 
attributed to better accounting of correlation effects in PRCC. To be more
precise, the importance of the correlation effects grows with increasing 
number of electrons, but, LPRCC theory is insufficient to incorporate 
the correlation effects as it considers only the linear terms. The PRCC 
theory, which includes the nonlinear terms, provides a better description
of the electron correlations.  This is borne by the fact that the PRCC 
results are in good agreement with the RCCSDT results, the difference 
between the two results is just $\approx$0.3\%. 

 Like in the previous cases, the contribution from the Breit interaction is 
small and the value is 0.008. Coming to the orbital energy correction arising
from VP, listed in Table. \ref{vp_ra}, there is a key difference from the 
other ions. The values of $\Delta\epsilon_{np_{1/2}}$, in addition to 
$\Delta\epsilon_{ns_{1/2}}$ are negative. 

\begin{table}[h]
    \caption{Contribution to $\alpha $ from different terms and their
             hermitian conjugates in the LPRCC and PRCC theory.}
    \label{result_lprcc}
    \begin{center}
    \begin{tabular}{lddddd}
        \hline
        Terms + h.c. & \multicolumn{1}{r}{$\rm{Mg}^{2+}$}
        & \multicolumn{1}{r}{$\rm{Ca}^{2+}$}
        & \multicolumn{1}{r}{$\rm{Sr}^{2+}$}
        & \multicolumn{1}{r}{$\rm{Ba}^{2+}$}
        & \multicolumn{1}{r}{$\rm{Ra}^{2+}$}  \\
        \hline
        & \multicolumn{5}{c}{LPRCC results}\\
        $\mathbf{T}_1^{(1)\dagger}\mathbf{D} $
        & 0.496 & 3.594 & 6.400 & 11.708 & 15.160 \\
        $\mathbf{T}_1{^{(1)\dagger}}\mathbf{D}T_2^{(0)} $
        & -0.008 & -0.180 & -0.330 & -0.676 & -0.864  \\
        $\mathbf{T}_1{^{(1)\dagger}}\mathbf{D}T_1^{(0)} $
        & 0.001 & -0.022 & -0.044 & -0.114 &   -0.108 \\
        $\mathbf{T}_2{^{(1)\dagger}}\mathbf{D}T_1^{(0)} $
        & -0.0001 & 0.004 & 0.008 & 0.020  &  0.018  \\
        $\mathbf{T}_2{^{(1)\dagger}}\mathbf{D}T_2^{(0)} $
        & 0.008  & 0.098 & 0.174  & 0.370  & 0.470   \\
        Normalization & 1.019 & 1.064 & 1.080 & 1.126 & 1.137  \\
        Total & 0.489 & 3.284  & 5.748 & 10.043 & 12.908 \\
         & \multicolumn{5}{c}{PRCC results}\\
        $\mathbf{T}_1^{(1)\dagger}\mathbf{D} $
        & 0.502   &  3.718  &  6.606  &  12.214  & 15.820  \\
        $\mathbf{T}_1{^{(1)\dagger}}\mathbf{D}T_2^{(0)} $
        & -0.008  & -0.188  & -0.344  & -0.710  & -0.908 \\
        $\mathbf{T}_2{^{(1)\dagger}}\mathbf{D}T_2^{(0)} $
        & 0.002   & -0.022  & -0.046  & -0.120  & -0.114 \\
        $\mathbf{T}_1{^{(1)\dagger}}\mathbf{D}T_1^{(0)} $
        & -0.0001 & -0.004  &  0.008  &  0.018  &  0.016 \\
        $\mathbf{T}_2{^{(1)\dagger}}\mathbf{D}T_1^{(0)} $
        & 0.008   &  0.092  &  0.162  &  0.338  &  0.424 \\
        Normalization & 1.019 & 1.064 &  1.080  & 1.126  & 1.137 \\
        Total & 0.495 & 3.387 & 5.913 &  10.426 & 13.402 \\
        \hline
    \end{tabular}
    \end{center}
\end{table}


\subsection{Core-polarization and pair correlation effects}
 In the previous sections we discussed the comparison between the results from
different theories, general trends and orbital energy corrections from VP.
To examine and investigate the contributions from various many-body effects,
which are encapsulated in different terms of LPRCC and PRCC, we isolate the
contributions from different terms through a series of computations. The
results are listed in Table. \ref{result_lprcc}. From the table it is evident 
that the leading term in the LPRCC as well as PRCC theory is 
$\{ \mathbf{T}_1^{(1)\dagger}\mathbf{D} +\rm{h.c}\}$. This is not surprising as
it is the term which subsumes the DF contribution and the RPA effects. Now to 
understand and quantify the RPA effects in these systems, we separate the 
core orbital contribution to $\alpha$. 
\begin{table}[h]
    \caption{Four leading contributions to
        $\{ \mathbf{T}_1^{(1)\dagger}\mathbf{D} + \rm{h.c} \}$ to $\alpha $
        in terms of the core spin-orbitals. }
    \label{result_t1d}
    \begin{center}
    \begin{tabular}{rrr}
        \hline
          \multicolumn{1}{c}{$\rm{Mg}^{2+}$} & 
          \multicolumn{1}{c}{$\rm{Ca}^{2+}$} &
          \multicolumn{1}{c}{$\rm{Sr}^{2+}$}  \\ \hline
        0.312 (2$p_{3/2}$) & 2.378 (3$p_{3/2}$) & 4.344 (4$p_{3/2}$) \\
        0.154 (2$p_{1/2}$) & 1.148 (3$p_{1/2}$) & 1.940 (4$p_{1/2}$) \\
        0.028 (2$s_{1/2}$) & 0.056 (3$s_{1/2}$) & 0.048 (4$s_{1/2}$) \\
        0.0002(1$s_{1/2}$) & 0.006 (2$p_{3/2}$) & 0.034 (3$d_{5/2}$) \\
        \hline
       \multicolumn{1}{c}{$\rm{Ba}^{2+}$}   & \multicolumn{1}{c}{$\rm{Ra}^{2+}$}
      &  \\  \hline
          8.182 (5$p_{3/2}$) & 11.766 (6$p_{3/2}$) & \\
          3.188 (5$p_{1/2}$) & 2.822  (6$p_{1/2}$) & \\
          0.162 (4$d_{5/2}$) & 0.338  (5$d_{5/2}$) & \\
          0.102 (4$d_{3/2}$) & 0.192  (5$d_{3/2}$) & \\ \hline
    \end{tabular}
    \end{center}
\end{table}
The four dominant contributions from the core orbitals to 
$\{ \mathbf{T}_1^{(1)\dagger}\mathbf{D} +\rm{h.c}\}$ are listed in table 
\ref{result_t1d}. For all the ions, the outermost $p_{3/2}$ orbital is the 
most dominant and this because of the larger radial extent of the $p_{3/2}$ 
orbitals. The next important contribution arises from the outermost 
$p_{1/2}$. A prominent feature that we observe in the results is the ratio
between the contribution from the outermost $p_{3/2}$ to the $p_{1/2}$ 
orbitals. The ratio are 2.03, 2.07, 2.24, 2.57 and 4.17 for $\rm{Mg}^{2+}$, 
$\rm{Ca}^{2+}$, $\rm{Sr}^{2+}$, $\rm{Ba}^{2+}$ and $\rm{Ra}^{2+}$,
respectively. The ratio increase with increasing $Z$ but
for $\rm{Ra}^{2+}$ it is 1.6 times higher than the $\rm{Ba}^{2+}$. This is
an important feature arising from the contraction of $p_{1/2}$ orbitals
due to the relativistic effects, which is more prominent in the heavier atoms
and ions. The third largest contribution arise from
$ns_{1/2}$ orbital in the case of $\rm{Mg}^{2+}$, $\rm{Ca}^{2+}$ and 
$\rm{Sr}^{2+}$. This is because the $ns_{1/2}$ orbital is energetically lower
than the $np_{1/2}$ and relativistic corrections are not large. However,
for $\rm{Ba}^{2+}$ and $\rm{Ra}^{2+}$, due to the relativistic contraction, the 
contribution from the outermost $ns_{1/2}$ is suppressed. And, the third 
largest contribution arises from the more diffused outer $nd_{5/2}$ orbital. 

   The next leading contribution arises from 
$\{\mathbf{T}_1{^{(1)\dagger}}\mathbf{D}T_2^{(0)} + \rm{h.c} \}$. The 
contribution from this term is much smaller and opposite in phase 
to the leading order term. A similar trend is observed in case of 
the noble gas atoms and was reported in one of our previous works 
\cite{chattopadhyay-12b}. Among the various terms the 
$\{\mathbf{T}_1{^{(1)\dagger}}\mathbf{D}T_1^{(0)} + \rm{h.c}\} $ has the
smallest contribution. This is because of the fact that 
$T_{1}^{(0)}$ and $\mathbf{T}_2{^{(1)}}$ have smaller amplitudes in the
RCC and PRCC theories, respectively. As can be seen from the table 
\ref{result_lprcc}, the overall contribution from the second order terms are 
0.0009, -0.100, -0.192, -0.400, -0.484 for $\rm{Mg}^{2+}$, $\rm{Ca}^{2+}$, 
$\rm{Sr}^{2+}$, $\rm{Ba}^{2+}$ and $\rm{Ra}^{2+}$, respectively. Except for 
$\rm{Mg}^{2+}$, the higher order terms gives a negative contribution to 
the $\alpha$.

To study the pair-correlation effects we examine the next to leading order
term, $\mathbf{T}_1{^{(1)\dagger}} \mathbf{D}T_2^{(0)}$ in more detail. In 
Table \ref{t1dt2_mg_ca}, \ref{t1dt2_sr_ba_ra} we list the four leading order 
core-orbital pairs which contribute to $\alpha$. The $(np_{3/2}, np_{3/2})$ 
orbital pairing gives the most dominant contribution. The next leading order 
contribution arises from the $(np_{3/2}, np_{1/2})$ orbital pairing. The same 
pattern is observed for all the doubly charged ions. For $\rm{Ra}^{2+}$ the 
fourth largest contribution arise from $(6p_{3/2}, 5d_{5/2})$ orbital pairing, 
but for other ions it is from $(np_{1/2}, np_{1/2})$ orbital pairing. This
is because of the relativistic effects, which contracts the outer $s_{1/2}$ 
orbital in $\rm{Ra}^{2+}$ more than the other ions.
One important point
to notice here is the higher order terms does not translate to higher accuracy 
as observed in the case of $\rm{Mg}^{2+}$ and $\rm{Ca}^{2+}$. 

\begin{table}[h]
  \caption{Core orbitals contribution from
           $\mathbf{T}_1{^{(1)\dagger}} \mathbf{D}T_2^{(0)}$ to $\alpha$
           of $\rm{Mg}^{2+}$ and $\rm{Ca}^{2+}$}
    \label{t1dt2_mg_ca}
    \begin{center}
    \begin{tabular}{dcdc}
       \hline
       \multicolumn{2}{c}{$\rm{Mg}^{2+}$}  & 
       \multicolumn{2}{c}{$\rm{Ca}^{2+}$} \\
       \hline
       -0.002  & $(2p_{3/2}, 2p_{3/2})$  &  -0.038 & $(3p_{3/2}, 3p_{3/2})$ \\
       -0.001  & $(2p_{3/2}, 2p_{1/2})$  &  -0.022 & $(3p_{3/2}, 3p_{1/2})$ \\
       -0.001  & $(2p_{1/2}, 2p_{3/2})$  &  -0.022 & $(3p_{1/2}, 3p_{3/2})$ \\
       -0.0004 & $(2p_{1/2}, 2p_{1/2})$  &  -0.009 & $(3p_{1/2}, 3p_{1/2})$ \\
     \hline
    \end{tabular}
    \end{center}
\end{table}

\begin{table}[h]
  \caption{Core orbitals contribution from
           $\mathbf{T}_1{^{(1)\dagger}} \mathbf{D}T_2^{(0)}$ to $\alpha$
           of $\rm{Sr}^{2+}$, $\rm{Ba}^{2+}$  and $\rm{Ra}^{2+}$}
    \label{t1dt2_sr_ba_ra}
    \begin{center}
    \begin{tabular}{dcdc}
       \hline
       \multicolumn{2}{c}{$\rm{Sr}^{2+}$}  & \multicolumn{2}{c}{$\rm{Ba}^{2+}$}
                    \\ \hline
       -0.069  & $(4p_{3/2}, 4p_{3/2})$  &  -0.132 & $(5p_{3/2}, 5p_{3/2})$\\
       -0.038  & $(4p_{3/2}, 4p_{1/2})$  &  -0.070 & $(5p_{3/2}, 5p_{1/2})$\\
       -0.036  & $(4p_{1/2}, 4p_{3/2})$  &  -0.061 & $(5p_{1/2}, 5p_{3/2})$\\
       -0.014  & $(4p_{1/2}, 4p_{1/2})$  &  -0.022 & $(5p_{1/2}, 5p_{1/2})$\\
     \hline
       \multicolumn{2}{c}{$\rm{Ra}^{2+}$} && \\ \hline
       -0.186 & $(6p_{3/2}, 6p_{3/2})$  && \\
       -0.077 & $(6p_{3/2}, 6p_{1/2})$  && \\
       -0.052 & $(6p_{1/2}, 6p_{3/2})$  && \\
       -0.039 & $(6p_{3/2}, 5d_{5/2})$  && \\ \hline
    \end{tabular}
    \end{center}
\end{table}


\subsection{Theoretical Uncertainty}

  We have isolated the following sources of uncertainty in the present
calculations. The first is the truncation of the numerical basis set.
We start our calculations with calculation with 9 symmetry and increase up
to 13 symmetry. Along with this we also increase the number of orbitals per
symmetry and we observe that our value of $\alpha$ converges for all the
doubly charged ions. So we can neglect this error safely. The second source of
error is associated with the truncation of RCC theory at the single and
doubles excitation in both the unperturbed and at the perturbed level. Based
on a series of test calculations, we estimate the contribution from triple 
excited cluster amplitudes to less than 0.2\% of the total value. So, we can
consider the upper bound on the uncertainty from the truncation of the RCC
and PRCC theories as 0.4\% for the heavier ions Sr$^{2+}$, Ba$^{2+}$ and
Ra$^{2+}$. Examining the trend in the results of Mg$^{2+}$ and Ca$^{2+}$, the 
uncertainty is likely to be higher for the PRCC resuls of these ions. But,
the LPRCC results could have an uncertainty less than $\approx$0.4\%. The 
third source of error is the truncation of the non-terminating series of 
$\alpha$. We terminate $e^{{\mathbf{T}^{(1)}}^\dagger}\mathbf{D}e^{T^{(0)}} 
+ e^{{T^{(0)}}^\dagger}\mathbf{D}e^{\mathbf{T}^{(1)}}$ at the second order in
cluster operator. However, based on our earlier study \cite{mani-10}, where 
we reported an iterative technique to calculate properties to all order, the 
contribution from the third and higher order terms is negligible. So, the 
uncertainty arising from the truncation in the expression of  $\alpha$ can be 
neglected. Quantum electrodynamic (QED) corrections is another source of 
uncertainty in the present calculation. We include the VP potential in the 
present work but the self-energy part of the radiative corrections is 
neglected. The self-energy correction is important for the heavy atoms 
\cite{mohr-98}. We can, however, safely neglect it from the error estimates 
as the contribution is less than the correction from Breit interaction, which 
accounts for at the most $0.1$\% of the total value. So, considering all the 
sources, the upper bound on the uncertainty of the present calculations is 
$\approx$0.4\% for the LPRCC results of Mg$^{2+}$ and Ca$^{2+}$, and PRCC 
results of Sr$^{2+}$, Ba$^{2+}$ and Ra$^{2+}$ ions.


\section{Conclusion}
  The electric dipole polarizability of doubly ionized alkaline-Earth-metal
ions calculated using the PRCC theory are in very good agreement with the  
previous theoretical results and experimental data. An important observation 
is, for the lighter ions Mg$^{2+}$ and Sr$^{2+}$ the inclusion of nonlinear 
terms in PRCC does not translate to better agreement with the experimental 
data. However, for the heavier ions, the nonlinear terms are essential to 
obtain results which are in agreement with the other results based on 
relativistic coupled-cluster theory. The correction from Breit interaction 
is show marginal increase with atomic number and this may be due to the 
radial dependence of the $\alpha$. 

  The changes in orbital energies, SCF and first order correction, with the 
VP potential reflects the short range nature of this potential. Further more,
there is an important change in the SCF energy correction $\Delta\epsilon$ 
with increasing $Z$. For lighter atoms only the $\Delta\epsilon$ of the core 
$ns_{1/2}$ are negative. But, for Ra$^{2+}$ in addition to the core $ns_{1/2}$,
the core $np_{1/2}$ orbitals also have negative $\Delta\epsilon$.

\vspace*{0.3cm}

\begin{acknowledgments}
We thank A. Roy and K. Suthar for useful discussions. The results presented in 
the paper are based on the computations using the 3TFLOP HPC Cluster at 
Physical Research Laboratory, Ahmedabad, India.
\end{acknowledgments}

\bibliography{references}{}

\begin{thebibliography}{42}%
\makeatletter
\providecommand \@ifxundefined [1]{%
 \@ifx{#1\undefined}
}%
\providecommand \@ifnum [1]{%
 \ifnum #1\expandafter \@firstoftwo
 \else \expandafter \@secondoftwo
 \fi
}%
\providecommand \@ifx [1]{%
 \ifx #1\expandafter \@firstoftwo
 \else \expandafter \@secondoftwo
 \fi
}%
\providecommand \natexlab [1]{#1}%
\providecommand \enquote  [1]{``#1''}%
\providecommand \bibnamefont  [1]{#1}%
\providecommand \bibfnamefont [1]{#1}%
\providecommand \citenamefont [1]{#1}%
\providecommand \href@noop [0]{\@secondoftwo}%
\providecommand \href [0]{\begingroup \@sanitize@url \@href}%
\providecommand \@href[1]{\@@startlink{#1}\@@href}%
\providecommand \@@href[1]{\endgroup#1\@@endlink}%
\providecommand \@sanitize@url [0]{\catcode `\\12\catcode `\$12\catcode
  `\&12\catcode `\#12\catcode `\^12\catcode `\_12\catcode `\%12\relax}%
\providecommand \@@startlink[1]{}%
\providecommand \@@endlink[0]{}%
\providecommand \url  [0]{\begingroup\@sanitize@url \@url }%
\providecommand \@url [1]{\endgroup\@href {#1}{\urlprefix }}%
\providecommand \urlprefix  [0]{URL }%
\providecommand \Eprint [0]{\href }%
\providecommand \doibase [0]{http://dx.doi.org/}%
\providecommand \selectlanguage [0]{\@gobble}%
\providecommand \bibinfo  [0]{\@secondoftwo}%
\providecommand \bibfield  [0]{\@secondoftwo}%
\providecommand \translation [1]{[#1]}%
\providecommand \BibitemOpen [0]{}%
\providecommand \bibitemStop [0]{}%
\providecommand \bibitemNoStop [0]{.\EOS\space}%
\providecommand \EOS [0]{\spacefactor3000\relax}%
\providecommand \BibitemShut  [1]{\csname bibitem#1\endcsname}%
\let\auto@bib@innerbib\@empty
\bibitem [{\citenamefont {Bonin}\ and\ \citenamefont
  {Kresin}(1997)}]{bonin-97}%
  \BibitemOpen
  \bibfield  {author} {\bibinfo {author} {\bibfnamefont {K.}~\bibnamefont
  {Bonin}}\ and\ \bibinfo {author} {\bibfnamefont {V.}~\bibnamefont {Kresin}},\
  }\href@noop {} {\emph {\bibinfo {title} {Electric-Dipole Polarizabilities of
  Atoms, Molecules and Clusters}}}\ (\bibinfo  {publisher} {World Scientific
  Publ.},\ \bibinfo {year} {1997})\BibitemShut {NoStop}%
\bibitem [{\citenamefont {Gould}\ and\ \citenamefont
  {Miller}(2005)}]{gould-05}%
  \BibitemOpen
  \bibfield  {author} {\bibinfo {author} {\bibfnamefont {H.}~\bibnamefont
  {Gould}}\ and\ \bibinfo {author} {\bibfnamefont {T.~M.}\ \bibnamefont
  {Miller}},\ }in\ \href {\doibase 10.1016/S1049-250X(05)51019-X} {\emph
  {\bibinfo {booktitle} {Advances In Atomic, Molecular, and Optical
  Physics}}},\ Vol.~\bibinfo {volume} {51},\ \bibinfo {editor} {edited by\
  \bibinfo {editor} {\bibfnamefont {H.}~\bibnamefont {Stroke}}}\ (\bibinfo
  {publisher} {Academic Press},\ \bibinfo {year} {2005})\ pp.\ \bibinfo {pages}
  {343--361}\BibitemShut {NoStop}%
\bibitem [{\citenamefont {Mitroy}\ \emph {et~al.}(2010)\citenamefont {Mitroy},
  \citenamefont {Safronova},\ and\ \citenamefont {Clark}}]{mitroy-10}%
  \BibitemOpen
  \bibfield  {author} {\bibinfo {author} {\bibfnamefont {J.}~\bibnamefont
  {Mitroy}}, \bibinfo {author} {\bibfnamefont {M.~S.}\ \bibnamefont
  {Safronova}}, \ and\ \bibinfo {author} {\bibfnamefont {C.~W.}\ \bibnamefont
  {Clark}},\ }\href {http://stacks.iop.org/0953-4075/43/i=20/a=202001}
  {\bibfield  {journal} {\bibinfo  {journal} {J. Phys. B}\ }\textbf {\bibinfo
  {volume} {43}},\ \bibinfo {pages} {202001} (\bibinfo {year}
  {2010})}\BibitemShut {NoStop}%
\bibitem [{\citenamefont {Coester}(1958)}]{coester-58}%
  \BibitemOpen
  \bibfield  {author} {\bibinfo {author} {\bibfnamefont {F.}~\bibnamefont
  {Coester}},\ }\href {\doibase 10.1016/0029-5582(58)90280-3} {\bibfield
  {journal} {\bibinfo  {journal} {Nucl. Phys.}\ }\textbf {\bibinfo {volume}
  {7}},\ \bibinfo {pages} {421 } (\bibinfo {year} {1958})}\BibitemShut
  {NoStop}%
\bibitem [{\citenamefont {Coester}\ and\ \citenamefont
  {K{\"{u}}mmel}(1960)}]{coester-60}%
  \BibitemOpen
  \bibfield  {author} {\bibinfo {author} {\bibfnamefont {F.}~\bibnamefont
  {Coester}}\ and\ \bibinfo {author} {\bibfnamefont {H.}~\bibnamefont
  {K{\"{u}}mmel}},\ }\href {\doibase 10.1016/0029-5582(60)90140-1} {\bibfield
  {journal} {\bibinfo  {journal} {Nucl. Phys.}\ }\textbf {\bibinfo {volume}
  {17}},\ \bibinfo {pages} {477 } (\bibinfo {year} {1960})}\BibitemShut
  {NoStop}%
\bibitem [{\citenamefont {Bartlett}\ and\ \citenamefont
  {Musia\l{}}(2007)}]{bartlett-07}%
  \BibitemOpen
  \bibfield  {author} {\bibinfo {author} {\bibfnamefont {R.~J.}\ \bibnamefont
  {Bartlett}}\ and\ \bibinfo {author} {\bibfnamefont {M.}~\bibnamefont
  {Musia\l{}}},\ }\href {\doibase 10.1103/RevModPhys.79.291} {\bibfield
  {journal} {\bibinfo  {journal} {Rev. Mod. Phys.}\ }\textbf {\bibinfo {volume}
  {79}},\ \bibinfo {pages} {291} (\bibinfo {year} {2007})}\BibitemShut
  {NoStop}%
\bibitem [{\citenamefont {Lindgren}\ and\ \citenamefont
  {Morrison}(1986)}]{lindgren-86}%
  \BibitemOpen
  \bibfield  {author} {\bibinfo {author} {\bibfnamefont {I.}~\bibnamefont
  {Lindgren}}\ and\ \bibinfo {author} {\bibfnamefont {J.}~\bibnamefont
  {Morrison}},\ }\href@noop {} {\emph {\bibinfo {title} {Atomic Many-Body
  Theory}}}\ (\bibinfo  {publisher} {Springer},\ \bibinfo {year} {2nd Edition,
  1986})\BibitemShut {NoStop}%
\bibitem [{\citenamefont {Shavitt}\ and\ \citenamefont
  {Bartlett}(2009)}]{shavitt-09}%
  \BibitemOpen
  \bibfield  {author} {\bibinfo {author} {\bibfnamefont {I.}~\bibnamefont
  {Shavitt}}\ and\ \bibinfo {author} {\bibfnamefont {R.}~\bibnamefont
  {Bartlett}},\ }\href@noop {} {\emph {\bibinfo {title} {Many-Body Methods in
  Chemistry and Physics: MBPT and Coupled-Cluster Theory}}}\ (\bibinfo
  {publisher} {Cambridge University Press},\ \bibinfo {year}
  {2009})\BibitemShut {NoStop}%
\bibitem [{\citenamefont {Mani}\ \emph {et~al.}(2009)\citenamefont {Mani},
  \citenamefont {Latha},\ and\ \citenamefont {Angom}}]{mani-09}%
  \BibitemOpen
  \bibfield  {author} {\bibinfo {author} {\bibfnamefont {B.~K.}\ \bibnamefont
  {Mani}}, \bibinfo {author} {\bibfnamefont {K.~V.~P.}\ \bibnamefont {Latha}},
  \ and\ \bibinfo {author} {\bibfnamefont {D.}~\bibnamefont {Angom}},\ }\href
  {\doibase 10.1103/PhysRevA.80.062505} {\bibfield  {journal} {\bibinfo
  {journal} {Phys. Rev. A}\ }\textbf {\bibinfo {volume} {80}},\ \bibinfo
  {pages} {062505} (\bibinfo {year} {2009})}\BibitemShut {NoStop}%
\bibitem [{\citenamefont {Nataraj}\ \emph {et~al.}(2008)\citenamefont
  {Nataraj}, \citenamefont {Sahoo}, \citenamefont {Das},\ and\ \citenamefont
  {Mukherjee}}]{nataraj-08}%
  \BibitemOpen
  \bibfield  {author} {\bibinfo {author} {\bibfnamefont {H.~S.}\ \bibnamefont
  {Nataraj}}, \bibinfo {author} {\bibfnamefont {B.~K.}\ \bibnamefont {Sahoo}},
  \bibinfo {author} {\bibfnamefont {B.~P.}\ \bibnamefont {Das}}, \ and\
  \bibinfo {author} {\bibfnamefont {D.}~\bibnamefont {Mukherjee}},\ }\href
  {\doibase 10.1103/PhysRevLett.101.033002} {\bibfield  {journal} {\bibinfo
  {journal} {Phys. Rev. Lett.}\ }\textbf {\bibinfo {volume} {101}},\ \bibinfo
  {pages} {033002} (\bibinfo {year} {2008})}\BibitemShut {NoStop}%
\bibitem [{\citenamefont {Pal}\ \emph {et~al.}(2007)\citenamefont {Pal},
  \citenamefont {Safronova}, \citenamefont {Johnson}, \citenamefont
  {Derevianko},\ and\ \citenamefont {Porsev}}]{pal-07}%
  \BibitemOpen
  \bibfield  {author} {\bibinfo {author} {\bibfnamefont {R.}~\bibnamefont
  {Pal}}, \bibinfo {author} {\bibfnamefont {M.~S.}\ \bibnamefont {Safronova}},
  \bibinfo {author} {\bibfnamefont {W.~R.}\ \bibnamefont {Johnson}}, \bibinfo
  {author} {\bibfnamefont {A.}~\bibnamefont {Derevianko}}, \ and\ \bibinfo
  {author} {\bibfnamefont {S.~G.}\ \bibnamefont {Porsev}},\ }\href {\doibase
  10.1103/PhysRevA.75.042515} {\bibfield  {journal} {\bibinfo  {journal} {Phys.
  Rev. A}\ }\textbf {\bibinfo {volume} {75}},\ \bibinfo {pages} {042515}
  (\bibinfo {year} {2007})}\BibitemShut {NoStop}%
\bibitem [{\citenamefont {Gopakumar}\ \emph {et~al.}(2001)\citenamefont
  {Gopakumar}, \citenamefont {Merlitz}, \citenamefont {Majumder}, \citenamefont
  {Chaudhuri}, \citenamefont {Das}, \citenamefont {Mahapatra},\ and\
  \citenamefont {Mukherjee}}]{geetha-01}%
  \BibitemOpen
  \bibfield  {author} {\bibinfo {author} {\bibfnamefont {G.}~\bibnamefont
  {Gopakumar}}, \bibinfo {author} {\bibfnamefont {H.}~\bibnamefont {Merlitz}},
  \bibinfo {author} {\bibfnamefont {S.}~\bibnamefont {Majumder}}, \bibinfo
  {author} {\bibfnamefont {R.~K.}\ \bibnamefont {Chaudhuri}}, \bibinfo {author}
  {\bibfnamefont {B.~P.}\ \bibnamefont {Das}}, \bibinfo {author} {\bibfnamefont
  {U.~S.}\ \bibnamefont {Mahapatra}}, \ and\ \bibinfo {author} {\bibfnamefont
  {D.}~\bibnamefont {Mukherjee}},\ }\href {\doibase 10.1103/PhysRevA.64.032502}
  {\bibfield  {journal} {\bibinfo  {journal} {Phys. Rev. A}\ }\textbf {\bibinfo
  {volume} {64}},\ \bibinfo {pages} {032502} (\bibinfo {year}
  {2001})}\BibitemShut {NoStop}%
\bibitem [{\citenamefont {Isaev}\ \emph {et~al.}(2004)\citenamefont {Isaev},
  \citenamefont {Petrov}, \citenamefont {Mosyagin}, \citenamefont {Titov},
  \citenamefont {Eliav},\ and\ \citenamefont {Kaldor}}]{isaev-04}%
  \BibitemOpen
  \bibfield  {author} {\bibinfo {author} {\bibfnamefont {T.~A.}\ \bibnamefont
  {Isaev}}, \bibinfo {author} {\bibfnamefont {A.~N.}\ \bibnamefont {Petrov}},
  \bibinfo {author} {\bibfnamefont {N.~S.}\ \bibnamefont {Mosyagin}}, \bibinfo
  {author} {\bibfnamefont {A.~V.}\ \bibnamefont {Titov}}, \bibinfo {author}
  {\bibfnamefont {E.}~\bibnamefont {Eliav}}, \ and\ \bibinfo {author}
  {\bibfnamefont {U.}~\bibnamefont {Kaldor}},\ }\href {\doibase
  10.1103/PhysRevA.69.030501} {\bibfield  {journal} {\bibinfo  {journal} {Phys.
  Rev. A}\ }\textbf {\bibinfo {volume} {69}},\ \bibinfo {pages} {030501}
  (\bibinfo {year} {2004})}\BibitemShut {NoStop}%
\bibitem [{\citenamefont {Hagen}\ \emph {et~al.}(2008)\citenamefont {Hagen},
  \citenamefont {Papenbrock}, \citenamefont {Dean},\ and\ \citenamefont
  {Hjorth-Jensen}}]{hagen-08}%
  \BibitemOpen
  \bibfield  {author} {\bibinfo {author} {\bibfnamefont {G.}~\bibnamefont
  {Hagen}}, \bibinfo {author} {\bibfnamefont {T.}~\bibnamefont {Papenbrock}},
  \bibinfo {author} {\bibfnamefont {D.~J.}\ \bibnamefont {Dean}}, \ and\
  \bibinfo {author} {\bibfnamefont {M.}~\bibnamefont {Hjorth-Jensen}},\ }\href
  {\doibase 10.1103/PhysRevLett.101.092502} {\bibfield  {journal} {\bibinfo
  {journal} {Phys. Rev. Lett.}\ }\textbf {\bibinfo {volume} {101}},\ \bibinfo
  {pages} {092502} (\bibinfo {year} {2008})}\BibitemShut {NoStop}%
\bibitem [{\citenamefont {Bishop}\ \emph {et~al.}(2009)\citenamefont {Bishop},
  \citenamefont {Li}, \citenamefont {Farnell},\ and\ \citenamefont
  {Campbell}}]{bishop-09}%
  \BibitemOpen
  \bibfield  {author} {\bibinfo {author} {\bibfnamefont {R.~F.}\ \bibnamefont
  {Bishop}}, \bibinfo {author} {\bibfnamefont {P.~H.~Y.}\ \bibnamefont {Li}},
  \bibinfo {author} {\bibfnamefont {D.~J.~J.}\ \bibnamefont {Farnell}}, \ and\
  \bibinfo {author} {\bibfnamefont {C.~E.}\ \bibnamefont {Campbell}},\ }\href
  {\doibase 10.1103/PhysRevB.79.174405} {\bibfield  {journal} {\bibinfo
  {journal} {Phys. Rev. B}\ }\textbf {\bibinfo {volume} {79}},\ \bibinfo
  {pages} {174405} (\bibinfo {year} {2009})}\BibitemShut {NoStop}%
\bibitem [{\citenamefont {Cohen}\ and\ \citenamefont
  {Roothaan}(1965)}]{cohen-65}%
  \BibitemOpen
  \bibfield  {author} {\bibinfo {author} {\bibfnamefont {H.~D.}\ \bibnamefont
  {Cohen}}\ and\ \bibinfo {author} {\bibfnamefont {C.~C.~J.}\ \bibnamefont
  {Roothaan}},\ }\href {\doibase 10.1063/1.1701512} {\bibfield  {journal}
  {\bibinfo  {journal} {J. Chem. Phys.}\ }\textbf {\bibinfo {volume} {43}},\
  \bibinfo {pages} {S34} (\bibinfo {year} {1965})}\BibitemShut {NoStop}%
\bibitem [{\citenamefont {Safronova}\ \emph {et~al.}(1999)\citenamefont
  {Safronova}, \citenamefont {Johnson},\ and\ \citenamefont
  {Derevianko}}]{safronova-99}%
  \BibitemOpen
  \bibfield  {author} {\bibinfo {author} {\bibfnamefont {M.~S.}\ \bibnamefont
  {Safronova}}, \bibinfo {author} {\bibfnamefont {W.~R.}\ \bibnamefont
  {Johnson}}, \ and\ \bibinfo {author} {\bibfnamefont {A.}~\bibnamefont
  {Derevianko}},\ }\href {\doibase 10.1103/PhysRevA.60.4476} {\bibfield
  {journal} {\bibinfo  {journal} {Phys. Rev. A}\ }\textbf {\bibinfo {volume}
  {60}},\ \bibinfo {pages} {4476} (\bibinfo {year} {1999})}\BibitemShut
  {NoStop}%
\bibitem [{\citenamefont {Derevianko}\ \emph {et~al.}(1999)\citenamefont
  {Derevianko}, \citenamefont {Johnson}, \citenamefont {Safronova},\ and\
  \citenamefont {Babb}}]{derevianko-99}%
  \BibitemOpen
  \bibfield  {author} {\bibinfo {author} {\bibfnamefont {A.}~\bibnamefont
  {Derevianko}}, \bibinfo {author} {\bibfnamefont {W.~R.}\ \bibnamefont
  {Johnson}}, \bibinfo {author} {\bibfnamefont {M.~S.}\ \bibnamefont
  {Safronova}}, \ and\ \bibinfo {author} {\bibfnamefont {J.~F.}\ \bibnamefont
  {Babb}},\ }\href {\doibase 10.1103/PhysRevLett.82.3589} {\bibfield  {journal}
  {\bibinfo  {journal} {Phys. Rev. Lett.}\ }\textbf {\bibinfo {volume} {82}},\
  \bibinfo {pages} {3589} (\bibinfo {year} {1999})}\BibitemShut {NoStop}%
\bibitem [{\citenamefont {Chattopadhyay}\ \emph
  {et~al.}(2012{\natexlab{a}})\citenamefont {Chattopadhyay}, \citenamefont
  {Mani},\ and\ \citenamefont {Angom}}]{chattopadhyay-12a}%
  \BibitemOpen
  \bibfield  {author} {\bibinfo {author} {\bibfnamefont {S.}~\bibnamefont
  {Chattopadhyay}}, \bibinfo {author} {\bibfnamefont {B.~K.}\ \bibnamefont
  {Mani}}, \ and\ \bibinfo {author} {\bibfnamefont {D.}~\bibnamefont {Angom}},\
  }\href {\doibase 10.1103/PhysRevA.86.022522} {\bibfield  {journal} {\bibinfo
  {journal} {Phys. Rev. A}\ }\textbf {\bibinfo {volume} {86}},\ \bibinfo
  {pages} {022522} (\bibinfo {year} {2012}{\natexlab{a}})}\BibitemShut
  {NoStop}%
\bibitem [{\citenamefont {Chattopadhyay}\ \emph
  {et~al.}(2012{\natexlab{b}})\citenamefont {Chattopadhyay}, \citenamefont
  {Mani},\ and\ \citenamefont {Angom}}]{chattopadhyay-12b}%
  \BibitemOpen
  \bibfield  {author} {\bibinfo {author} {\bibfnamefont {S.}~\bibnamefont
  {Chattopadhyay}}, \bibinfo {author} {\bibfnamefont {B.~K.}\ \bibnamefont
  {Mani}}, \ and\ \bibinfo {author} {\bibfnamefont {D.}~\bibnamefont {Angom}},\
  }\href {\doibase 10.1103/PhysRevA.86.062508} {\bibfield  {journal} {\bibinfo
  {journal} {Phys. Rev. A}\ }\textbf {\bibinfo {volume} {86}},\ \bibinfo
  {pages} {062508} (\bibinfo {year} {2012}{\natexlab{b}})}\BibitemShut
  {NoStop}%
\bibitem [{\citenamefont {Chattopadhyay}\ \emph {et~al.}(2013)\citenamefont
  {Chattopadhyay}, \citenamefont {Mani},\ and\ \citenamefont
  {Angom}}]{chattopadhyay-13}%
  \BibitemOpen
  \bibfield  {author} {\bibinfo {author} {\bibfnamefont {S.}~\bibnamefont
  {Chattopadhyay}}, \bibinfo {author} {\bibfnamefont {B.~K.}\ \bibnamefont
  {Mani}}, \ and\ \bibinfo {author} {\bibfnamefont {D.}~\bibnamefont {Angom}},\
  }\href {\doibase 10.1103/PhysRevA.87.042520} {\bibfield  {journal} {\bibinfo
  {journal} {Phys. Rev. A}\ }\textbf {\bibinfo {volume} {87}},\ \bibinfo
  {pages} {042520} (\bibinfo {year} {2013})}\BibitemShut {NoStop}%
\bibitem [{\citenamefont {Douglas}\ and\ \citenamefont
  {Kroll}(1974)}]{douglas-74}%
  \BibitemOpen
  \bibfield  {author} {\bibinfo {author} {\bibfnamefont {M.}~\bibnamefont
  {Douglas}}\ and\ \bibinfo {author} {\bibfnamefont {N.~M.}\ \bibnamefont
  {Kroll}},\ }\href {\doibase 10.1016/0003-4916(74)90333-9} {\bibfield
  {journal} {\bibinfo  {journal} {Ann. Phys.}\ }\textbf {\bibinfo {volume}
  {82}},\ \bibinfo {pages} {89} (\bibinfo {year} {1974})}\BibitemShut {NoStop}%
\bibitem [{\citenamefont {Lim}\ and\ \citenamefont
  {Schwerdtfeger}(2004)}]{lim-04}%
  \BibitemOpen
  \bibfield  {author} {\bibinfo {author} {\bibfnamefont {I.~S.}\ \bibnamefont
  {Lim}}\ and\ \bibinfo {author} {\bibfnamefont {P.}~\bibnamefont
  {Schwerdtfeger}},\ }\href {\doibase 10.1103/PhysRevA.70.062501} {\bibfield
  {journal} {\bibinfo  {journal} {Phys. Rev. A}\ }\textbf {\bibinfo {volume}
  {70}},\ \bibinfo {pages} {062501} (\bibinfo {year} {2004})}\BibitemShut
  {NoStop}%
\bibitem [{\citenamefont {Mohr}\ \emph {et~al.}(2012)\citenamefont {Mohr},
  \citenamefont {Taylor},\ and\ \citenamefont {Newell}}]{mohr-12}%
  \BibitemOpen
  \bibfield  {author} {\bibinfo {author} {\bibfnamefont {P.~J.}\ \bibnamefont
  {Mohr}}, \bibinfo {author} {\bibfnamefont {B.~N.}\ \bibnamefont {Taylor}}, \
  and\ \bibinfo {author} {\bibfnamefont {D.~B.}\ \bibnamefont {Newell}},\
  }\href {\doibase 10.1103/RevModPhys.84.1527} {\bibfield  {journal} {\bibinfo
  {journal} {Rev. Mod. Phys.}\ }\textbf {\bibinfo {volume} {84}},\ \bibinfo
  {pages} {1527} (\bibinfo {year} {2012})}\BibitemShut {NoStop}%
\bibitem [{\citenamefont {Brown}\ and\ \citenamefont
  {Ravenhall}(1951)}]{brown-51}%
  \BibitemOpen
  \bibfield  {author} {\bibinfo {author} {\bibfnamefont {G.~E.}\ \bibnamefont
  {Brown}}\ and\ \bibinfo {author} {\bibfnamefont {D.~G.}\ \bibnamefont
  {Ravenhall}},\ }\href {\doibase 10.1098/rspa.1951.0181} {\bibfield  {journal}
  {\bibinfo  {journal} {Proceedings of the Royal Society of London. Series A.
  Mathematical and Physical Sciences}\ }\textbf {\bibinfo {volume} {208}},\
  \bibinfo {pages} {552} (\bibinfo {year} {1951})}\BibitemShut {NoStop}%
\bibitem [{\citenamefont {Sucher}(1980)}]{sucher-80}%
  \BibitemOpen
  \bibfield  {author} {\bibinfo {author} {\bibfnamefont {J.}~\bibnamefont
  {Sucher}},\ }\href {\doibase 10.1103/PhysRevA.22.348} {\bibfield  {journal}
  {\bibinfo  {journal} {Phys. Rev. A}\ }\textbf {\bibinfo {volume} {22}},\
  \bibinfo {pages} {348} (\bibinfo {year} {1980})}\BibitemShut {NoStop}%
\bibitem [{\citenamefont {Stanton}\ and\ \citenamefont
  {Havriliak}(1984)}]{stanton-84}%
  \BibitemOpen
  \bibfield  {author} {\bibinfo {author} {\bibfnamefont {R.~E.}\ \bibnamefont
  {Stanton}}\ and\ \bibinfo {author} {\bibfnamefont {S.}~\bibnamefont
  {Havriliak}},\ }\href {\doibase 10.1063/1.447865} {\bibfield  {journal}
  {\bibinfo  {journal} {J. Chem. Phys.}\ }\textbf {\bibinfo {volume} {81}},\
  \bibinfo {pages} {1910} (\bibinfo {year} {1984})}\BibitemShut {NoStop}%
\bibitem [{\citenamefont {Mohanty}\ and\ \citenamefont
  {Clementi}(1990)}]{mohanty-90}%
  \BibitemOpen
  \bibfield  {author} {\bibinfo {author} {\bibfnamefont {A.~K.}\ \bibnamefont
  {Mohanty}}\ and\ \bibinfo {author} {\bibfnamefont {E.}~\bibnamefont
  {Clementi}},\ }\href {\doibase 10.1063/1.459110} {\bibfield  {journal}
  {\bibinfo  {journal} {J. Chem. Phys.}\ }\textbf {\bibinfo {volume} {93}},\
  \bibinfo {pages} {1829} (\bibinfo {year} {1990})}\BibitemShut {NoStop}%
\bibitem [{\citenamefont {Grant}(2006)}]{grant-06}%
  \BibitemOpen
  \bibfield  {author} {\bibinfo {author} {\bibfnamefont {I.}~\bibnamefont
  {Grant}},\ }in\ \href {\doibase 10.1007/978-0-387-26308-3_22} {\emph
  {\bibinfo {booktitle} {Springer Handbook of Atomic, Molecular, and Optical
  Physics}}},\ \bibinfo {editor} {edited by\ \bibinfo {editor} {\bibfnamefont
  {G.}~\bibnamefont {Drake}}}\ (\bibinfo  {publisher} {Springer New York},\
  \bibinfo {year} {2006})\ pp.\ \bibinfo {pages} {325--357}\BibitemShut
  {NoStop}%
\bibitem [{\citenamefont {Grant}(2010)}]{grant-10}%
  \BibitemOpen
  \bibfield  {author} {\bibinfo {author} {\bibfnamefont {I.~P.}\ \bibnamefont
  {Grant}},\ }\href@noop {} {\emph {\bibinfo {title} {Relativistic Quantum
  Theory of Atoms and Molecules: Theory and Computation}}}\ (\bibinfo
  {publisher} {Springer},\ \bibinfo {year} {2010})\BibitemShut {NoStop}%
\bibitem [{\citenamefont {Uehling}(1935)}]{uehling-35}%
  \BibitemOpen
  \bibfield  {author} {\bibinfo {author} {\bibfnamefont {E.~A.}\ \bibnamefont
  {Uehling}},\ }\href {\doibase 10.1103/PhysRev.48.55} {\bibfield  {journal}
  {\bibinfo  {journal} {Phys. Rev.}\ }\textbf {\bibinfo {volume} {48}},\
  \bibinfo {pages} {55} (\bibinfo {year} {1935})}\BibitemShut {NoStop}%
\bibitem [{\citenamefont {Parpia}\ and\ \citenamefont
  {Mohanty}(1992)}]{parpia-92a}%
  \BibitemOpen
  \bibfield  {author} {\bibinfo {author} {\bibfnamefont {F.~A.}\ \bibnamefont
  {Parpia}}\ and\ \bibinfo {author} {\bibfnamefont {A.~K.}\ \bibnamefont
  {Mohanty}},\ }\href {\doibase 10.1103/PhysRevA.46.3735} {\bibfield  {journal}
  {\bibinfo  {journal} {Phys. Rev. A}\ }\textbf {\bibinfo {volume} {46}},\
  \bibinfo {pages} {3735} (\bibinfo {year} {1992})}\BibitemShut {NoStop}%
\bibitem [{\citenamefont {Fullerton}\ and\ \citenamefont
  {Rinker}(1976)}]{fullerton-76}%
  \BibitemOpen
  \bibfield  {author} {\bibinfo {author} {\bibfnamefont {L.~W.}\ \bibnamefont
  {Fullerton}}\ and\ \bibinfo {author} {\bibfnamefont {G.~A.}\ \bibnamefont
  {Rinker}},\ }\href {\doibase 10.1103/PhysRevA.13.1283} {\bibfield  {journal}
  {\bibinfo  {journal} {Phys. Rev. A}\ }\textbf {\bibinfo {volume} {13}},\
  \bibinfo {pages} {1283} (\bibinfo {year} {1976})}\BibitemShut {NoStop}%
\bibitem [{\citenamefont {Johnson}\ \emph {et~al.}(2001)\citenamefont
  {Johnson}, \citenamefont {Bednyakov},\ and\ \citenamefont
  {Soff}}]{johnson-01}%
  \BibitemOpen
  \bibfield  {author} {\bibinfo {author} {\bibfnamefont {W.~R.}\ \bibnamefont
  {Johnson}}, \bibinfo {author} {\bibfnamefont {I.}~\bibnamefont {Bednyakov}},
  \ and\ \bibinfo {author} {\bibfnamefont {G.}~\bibnamefont {Soff}},\ }\href
  {\doibase 10.1103/PhysRevLett.87.233001} {\bibfield  {journal} {\bibinfo
  {journal} {Phys. Rev. Lett.}\ }\textbf {\bibinfo {volume} {87}},\ \bibinfo
  {pages} {233001} (\bibinfo {year} {2001})}\BibitemShut {NoStop}%
\bibitem [{\citenamefont {Chaudhuri}\ \emph {et~al.}(1999)\citenamefont
  {Chaudhuri}, \citenamefont {Panda},\ and\ \citenamefont
  {Das}}]{chaudhuri-99}%
  \BibitemOpen
  \bibfield  {author} {\bibinfo {author} {\bibfnamefont {R.~K.}\ \bibnamefont
  {Chaudhuri}}, \bibinfo {author} {\bibfnamefont {P.~K.}\ \bibnamefont
  {Panda}}, \ and\ \bibinfo {author} {\bibfnamefont {B.~P.}\ \bibnamefont
  {Das}},\ }\href {\doibase 10.1103/PhysRevA.59.1187} {\bibfield  {journal}
  {\bibinfo  {journal} {Phys. Rev. A}\ }\textbf {\bibinfo {volume} {59}},\
  \bibinfo {pages} {1187} (\bibinfo {year} {1999})}\BibitemShut {NoStop}%
\bibitem [{\citenamefont {Parpia}\ \emph {et~al.}(1996)\citenamefont {Parpia},
  \citenamefont {Froese~Fischer},\ and\ \citenamefont {Grant}}]{parpia-96}%
  \BibitemOpen
  \bibfield  {author} {\bibinfo {author} {\bibfnamefont {F.~A.}\ \bibnamefont
  {Parpia}}, \bibinfo {author} {\bibfnamefont {C.}~\bibnamefont
  {Froese~Fischer}}, \ and\ \bibinfo {author} {\bibfnamefont {I.~P.}\
  \bibnamefont {Grant}},\ }\href {\doibase 10.1016/0010-4655(95)00136-0}
  {\bibfield  {journal} {\bibinfo  {journal} {Comp. Phys. Comm.}\ }\textbf
  {\bibinfo {volume} {94}},\ \bibinfo {pages} {249 } (\bibinfo {year}
  {1996})}\BibitemShut {NoStop}%
\bibitem [{\citenamefont {Pulay}(1980)}]{pulay-80}%
  \BibitemOpen
  \bibfield  {author} {\bibinfo {author} {\bibfnamefont {P.}~\bibnamefont
  {Pulay}},\ }\href {\doibase 10.1016/0009-2614(80)80396-4} {\bibfield
  {journal} {\bibinfo  {journal} {Chem. Phys. Lett.}\ }\textbf {\bibinfo
  {volume} {73}},\ \bibinfo {pages} {393 } (\bibinfo {year}
  {1980})}\BibitemShut {NoStop}%
\bibitem [{\citenamefont {Derevianko}\ \emph {et~al.}(2004)\citenamefont
  {Derevianko}, \citenamefont {Ravaine},\ and\ \citenamefont
  {Johnson}}]{derevianko-04}%
  \BibitemOpen
  \bibfield  {author} {\bibinfo {author} {\bibfnamefont {A.}~\bibnamefont
  {Derevianko}}, \bibinfo {author} {\bibfnamefont {B.}~\bibnamefont {Ravaine}},
  \ and\ \bibinfo {author} {\bibfnamefont {W.~R.}\ \bibnamefont {Johnson}},\
  }\href {\doibase 10.1103/PhysRevA.69.054502} {\bibfield  {journal} {\bibinfo
  {journal} {Phys. Rev. A}\ }\textbf {\bibinfo {volume} {69}},\ \bibinfo
  {pages} {054502} (\bibinfo {year} {2004})}\BibitemShut {NoStop}%
\bibitem [{\citenamefont {Johnson}\ \emph {et~al.}(1983)\citenamefont
  {Johnson}, \citenamefont {Kolb},\ and\ \citenamefont {Huang}}]{johnson-83}%
  \BibitemOpen
  \bibfield  {author} {\bibinfo {author} {\bibfnamefont {W.}~\bibnamefont
  {Johnson}}, \bibinfo {author} {\bibfnamefont {D.}~\bibnamefont {Kolb}}, \
  and\ \bibinfo {author} {\bibfnamefont {K.-N.}\ \bibnamefont {Huang}},\ }\href
  {\doibase 10.1016/0092-640X(83)90020-7} {\bibfield  {journal} {\bibinfo
  {journal} {At. Data and Nucl. Data Tables}\ }\textbf {\bibinfo {volume}
  {28}},\ \bibinfo {pages} {333 } (\bibinfo {year} {1983})}\BibitemShut
  {NoStop}%
\bibitem [{\citenamefont {{\"O}pik}(1967)}]{opik-67}%
  \BibitemOpen
  \bibfield  {author} {\bibinfo {author} {\bibfnamefont {U.}~\bibnamefont
  {{\"O}pik}},\ }\href {http://stacks.iop.org/0370-1328/92/i=3/a=308}
  {\bibfield  {journal} {\bibinfo  {journal} {Proceedings of the Physical
  Society}\ }\textbf {\bibinfo {volume} {92}},\ \bibinfo {pages} {566}
  (\bibinfo {year} {1967})}\BibitemShut {NoStop}%
\bibitem [{\citenamefont {Mani}\ and\ \citenamefont {Angom}(2010)}]{mani-10}%
  \BibitemOpen
  \bibfield  {author} {\bibinfo {author} {\bibfnamefont {B.~K.}\ \bibnamefont
  {Mani}}\ and\ \bibinfo {author} {\bibfnamefont {D.}~\bibnamefont {Angom}},\
  }\href {\doibase 10.1103/PhysRevA.81.042514} {\bibfield  {journal} {\bibinfo
  {journal} {Phys. Rev. A}\ }\textbf {\bibinfo {volume} {81}},\ \bibinfo
  {pages} {042514} (\bibinfo {year} {2010})}\BibitemShut {NoStop}%
\bibitem [{\citenamefont {Mohr}\ \emph {et~al.}(1998)\citenamefont {Mohr},
  \citenamefont {Plunien},\ and\ \citenamefont {Soff}}]{mohr-98}%
  \BibitemOpen
  \bibfield  {author} {\bibinfo {author} {\bibfnamefont {P.~J.}\ \bibnamefont
  {Mohr}}, \bibinfo {author} {\bibfnamefont {G.}~\bibnamefont {Plunien}}, \
  and\ \bibinfo {author} {\bibfnamefont {G.}~\bibnamefont {Soff}},\ }\href
  {\doibase 10.1016/S0370-1573(97)00046-X} {\bibfield  {journal} {\bibinfo
  {journal} {Phys. Rep}\ }\textbf {\bibinfo {volume} {293}},\ \bibinfo {pages}
  {227 } (\bibinfo {year} {1998})}\BibitemShut {NoStop}%
\end{thebibliography}%
\bibliographystyle{apsrev4-1}

\end{document}